\documentclass[a4paper,10pt]{article}
\usepackage{bm,amsmath,amssymb,mathrsfs,graphicx,citesort,enumerate}
\newcommand{\nc}{\newcommand}
\nc{\rnc}{\renewcommand}
\nc{\beq}{\begin{equation}}
\nc{\eeq}{\end{equation}}
\nc{\nn}{\nonumber}
\rnc{\(}{\left(}
\rnc{\)}{\right)}
\rnc{\[}{\left[}
\rnc{\]}{\right]}

\newtheorem{definition}{Definition}
\newtheorem{theorem}{Theorem}
\newtheorem{proposition}{Proposition}
\newtheorem{lemma}{Lemma}
\newtheorem{corollary}{Corollary}

\numberwithin{definition}{section}
\numberwithin{equation}{section}
\numberwithin{lemma}{section}
\numberwithin{proposition}{section}
\numberwithin{theorem}{section}
\numberwithin{corollary}{section}

\nc{\sq}{\qquad $\blacksquare$}
\nc{\wsq}{\qquad $\square$}

\begin{document}%
%
\title{Reflection equation for the $N=3$ Cremmer-Gervais $R$-matrix}
\author{Kohei Motegi$^{1}$\thanks{E-mail: motegi@gokutan.c.u-tokyo.ac.jp} \,
and  Yuji Yamada$^{2}$\thanks{E-mail: yamada@rkmath.rikkyo.ac.jp}\\\\
\it $^{1}$Institute of physics, University of Tokyo, \\
\it Komaba 3-8-1,
Meguro-ku, Tokyo 153-8902, Japan, \\
\it $^{2}$ Department of Mathematics, Rikkyo University, \\
\it Nishi Ikebukuro 3-34-1,
Toshima-ku, Tokyo 171-8501, Japan \\\\
\\}

\date{\today}
 
 
\maketitle

%
%
\begin{abstract}
We consider the reflection equation of the
$N$=3 Cremmer-Gervais $R$-matrix.
The reflection equation is shown to be equivalent to
38 equations which do not depend on the parameter of the
$R$-matrix, $q$.
Solving those 38 equations,
the solution space is found to be the union of two types of spaces,
each of which is parametrized by the algebraic variety
$\mathbb{P}^1(\mathbb{C}) \times
\mathbb{P}^1(\mathbb{C}) \times
\mathbb{P}^2(\mathbb{C})$
and
$
\mathbb{C} \times
\mathbb{P}^1(\mathbb{C}) \times
\mathbb{P}^2(\mathbb{C})$.
\end{abstract}

%
\section{Introduction}
The Yang-Baxter equation is the sufficient condition for the
integrability of the one-dimensional quantum systems (or
the two-dimensional classical statistical systems), i.e.,
it ensures the commutativity of the transfer matrices.
Based on the Yang-Baxter equation, the quantum inverse scattering method
was developed, which enables us to calculate bulk quantities and correlation
functions.

Under the open boundary condition, besides the Yang-Baxter equation,
the reflection equation guarantees the existence of the commutative 
family of transfer matrices \cite{Sklyanin}.

Up to now, there are many works about the reflection equation 
\cite{Sklyanin,KS,DG,BFKZ,IOZ,LS1,LS2,Y2,N,BS,DN,BB,Y1,AACDFG,LM,DK,DFIL}.
Taking the XXZ chain for example, the diagonal solution was obtained in
\cite{Sklyanin,KS}, the general solution in \cite{DG}.
For $N \ge 3$-state models, most of the solutions are obtained by imposing
initial conditions. There is also an approach from the quantum group 
\cite{N,BS,DN,BB}.
In \cite{N}, some nondiagonal solutions of the refelction equation for the
$U_q(\widehat{sl_2})$ $R$-matrix were obtained from the intertwining condition.
Moreover, a complete description
in terms of current algebra has been accomplished in \cite{BS}.

In this paper, we consider the reflection equation of the $N=3$ Cremmer-Gervais 
$R$-matrix. This $R$-matrix originally appeared in the context of the
Toda field theory \cite{CG} as a constant $R$-matrix.
Recently, the Baxterized $R$-matrix was derived \cite{EH} by
taking approriate trigonometric degeneration of the
Shibukawa-Ueno $R$-operator \cite{SU}.
We determine the full solution space without imposing
any condition such as the initial condition.

The main result is that we found the solution
space is the union of two types of spaces, which are parametzied by
algebraic varieties.
The first type is parametrized by
$\mathbb{P}^1(\mathbb{C}) \times
\mathbb{P}^1(\mathbb{C}) \times
\mathbb{P}^2(\mathbb{C})$, and the solution
can be explicitly expressed as
\begin{align}
K_{\mathrm{I}}(z,(B_1, B_2) \times (D_1, D_2) \times (E_1, E_2, E_3))=&
K_{\mathrm{I},0}(z,D_1,D_2,E_1)-z^6 T K_{\mathrm{I},0}(z^{-1},D_2,D_1,E_3)T \nn \\
&+K_{\mathrm{I},1}(z,B_1,B_2,D_1,E_2)-z^6 T K_{\mathrm{I},1}(z^{-1},B_2,B_1,D_2,E_2)T, \nn 
\end{align}
where $(B_1, B_2) \times (D_1, D_2) \times (E_1, E_2, E_3) \in
\mathbb{P}^1(\mathbb{C}) \times
\mathbb{P}^1(\mathbb{C}) \times
\mathbb{P}^2(\mathbb{C})$ and
\begin{align}
K_{\mathrm{I},0}(z,D_1,D_2,E_1)
=&E_1^2 \left(
\begin{array}{ccc}
D_2^2z^2  & D_1 D_2(z^4-1) &D_1^2  z^2(z^4-1) \\
0 &D_2^2 z^2 & D_1 D_2(z^4-1) \\
0 & 0 & D_2^2  z^2
\end{array}
\right), \nn \\
K_{\mathrm{I},1}(z,B_1,B_2,D_1,E_2)
=&-D_1 E_2 z^2 \left(
\begin{array}{ccc}
B_1 & 0 & 0 \\
0 & B_1 & B_2(1-z^4) \\
0 & 0 & B_1 z^4
\end{array}
\right), \nn \\
T=& \left(
\begin{array}{ccc}
0 & 0 & 1 \\
0 & 1 & 0 \\
1 & 0 & 0
\end{array}
\right). \nn
\end{align}
The second type is parametrized by $
\mathbb{C} \times
\mathbb{P}^1(\mathbb{C}) \times
\mathbb{P}^2(\mathbb{C})$,
and the explicit expression is
\begin{align}
&K_{\mathrm{II}}(z, (b) \times (F_1, F_2) \times (G_1, G_2, G_3))
\nn \\
&=bz^2 \mathrm{Id}+ K_{\mathrm{II},0}(z,F_1,G_1,G_2,G_3)
-z^4 T K_{\mathrm{II},0}(z^{-1},F_2,G_3,-G_2,G_1)T, \nn
\end{align}
where $(b) \times (F_1, F_2) \times (G_1, G_2, G_3) \in 
\mathbb{C} \times
\mathbb{P}^1(\mathbb{C}) \times
\mathbb{P}^2(\mathbb{C})$ and
\begin{align}
K_{\mathrm{II},0}(z,F_1,G_1,G_2,G_3)
=&-F_1 \left(
\begin{array}{ccc}
G_3  & 0 & G_1(1-z^4) \\
0 &G_3 & G_2(1-z^4) \\
0 & 0 & G_3 z^4
\end{array}
\right). \nn
\end{align}

In the next section, we define the Cremmer-Gervais $R$-matrix
and state again the main theorem, which is about the full solution 
space of the reflection equation.
We also derive some properties of the reflection equation
coming from the symmetries of the Cremmer-Gervais $R$-matrix,
which we use for the proof of the main theorem Th \ref{maintheorem},
given in sections 3 and 4.
Section 5 is devoted to the conclusion.
\section{The $N$=3 Cremmer-Gervais $R$-matrix and the reflection equation}
\subsection{The Cremmer-Gervias $R$-matrix}
We denote the standard orthonormal basis of $\mathbb{C}^3$ by
$\{ e_{0}, e_{1}, e_{2} \}$.
The matrix element $A_{j}^{i}$ of
$A \in \textrm{End}(\mathbb{C}^3)$
with respect to this basis is defined as
\begin{equation}
A e_{j}=\sum_{i=0}^{2}e_{i} A_{j}^{i}. \nonumber
\end{equation}
We also define $G$ and $T$ as
\begin{equation}
G e_{j}=\omega^j e_{j}, \ \ \ \ \ Te_{j}=e_{2-j}, \label{GT}
\end{equation}
where $\omega^3=1$. \\
The original Cremmer-Gervais $R$-matrix has two parameters
besides the spectral parameter \cite{CG,EH}. As a solution to
the Yang-Baxter equation, it is equivalent to the
$R$-matrix with one spectral parameter and
one nonspectral parameter which is defined below.
\begin{definition} $\mathrm{\cite{CG,EH}}$
The $N=3$ Cremmer-Gervais $R$-matrix
$R^{CG}(z,q) \in \mathrm{End}( \mathbb{C}^3 \otimes 
\mathbb{C}^3)$ is defined as
$ \[ R^{CG}(z,q) \]_{kl}^{ij}  $
\begin{eqnarray}
= 
 \left\{
\begin{array}{cc}
(q z^{-1}-q^{-1} z) \slash (q-q^{-1})(z-z^{-1}), & \mathrm{for} \ i=j=k=l, \\
-q^{\mathrm{sgn}(k-l)} \slash (q-q^{-1}), & \mathrm{for} \ i=k \neq  j=l, \\
z^{\mathrm{sgn}(l-k)} \slash (z-z^{-1}), & \mathrm{for} \ l=i \neq  k=j, \\
\mathrm{sgn}(l-k), & \mathrm{for} \ \mathrm{min}(k,l) < i < \mathrm{max}(k,l), \
i+j=k+l, \\
0, & \mathrm{otherwise}. \label{cremmer-gervais}
\end{array}
\right.
\end{eqnarray}
 \hspace*{\fill}  \sq
\end{definition}
\begin{theorem}
The Cremmer-Gervais $R$-matrix $R^{CG}(z,q)$ satisfies the Yang-Baxter equation
\begin{equation}
R_{12}(z_{1})R_{13}(z_{1} z_{2})R_{23}(z_{2})=
R_{23}(z_{2})R_{13}(z_{1} z_{2})R_{12}(z_{1})
\in \mathrm{End}(\mathbb{C}^3 \otimes \mathbb{C}^3 \otimes
\mathbb{C}^3). \label{yang-baxter}
\end{equation}
 \hspace*{\fill}    \sq
\end{theorem}
We can immediately see from the defintion that the
Cremmer-Gervais $R$-matrix $R^{CG}(z,q)$ has the following properties.
\begin{proposition}
The Cremmer-Gervais $R$-matrix $R^{CG}(z,q)$ has the following properties.
\begin{align}
Unitarity:  & \ R_{12}^{CG}(z)R_{21}^{CG}(z^{-1})=
\frac{(q^2-z^2)(1-q^2 z^2)}{(q^2-1)^2 (z^2-1)^2}\mathrm{Id}, 
\label{unitarity} \\
Conservation \ law: & \ R^{CG}(z)(G \otimes G)=(G \otimes G)R^{CG}(z),
\label{conservation} \\
T\textrm{-}invariance: & \ R^{CG}(z,q)=
-(T \otimes T)R^{CG}(z^{-1}, q^{-1})(T \otimes T), \label{t-invariance}
\end{align}
where 
$R_{21}(z)=PR_{12}(z)P, P(x \otimes y)=y \otimes x,$ for any $x, y \in
\mathbb{C}^3$.       
\hspace*{\fill}     \sq
\end{proposition}  
\eqref{unitarity} is used to show that
$R^{CG}(z,q)$ satisfies the Yang-Baxter equation \eqref{yang-baxter}.
\eqref{conservation} means that
$\[ R^{CG}(z) \]_{kl}^{ij} \equiv 0$ unless $i+j=k+l \pmod 3$,
which the Belavin $R$-matrix also satisfies. 
On the other hand, \eqref{t-invariance} is the symmetry peculiar to the 
Cremmer-Gervais $R$-matrix.
\begin{definition}
The reflection equation is
\begin{eqnarray}
R_{12}(z_{1}/z_{2})K_{1}(z_{1})R_{21}(z_{1} z_{2})K_{2}(z_{2})
=K_{2}(z_{2})R_{12}(z_{1} z_{2})K_{1}(z_{1})R_{21}(z_{1}/ z_{2})
\in \mathrm{End}(\mathbb{C}^3 \otimes \mathbb{C}^3), \label{re}
\end{eqnarray}
where $R_{12}(z)=R(z)$ is the Cremmer-Gervais $R$-matrix 
\eqref{cremmer-gervais}, and \\
$K_{1}(z)=K(z) \otimes \mathrm{Id},
K_{2}(z)=\mathrm{Id} \otimes K(z)$.   \hspace*{\fill}  \sq
\end{definition}
We obtained all the meromorphic solutions to the reflection equation
of the
Cremmer-Gervais $R$-matrix $R^{CG}(z,q)$ which are not identically zero,
i.e, we determined the full solution space
\begin{align}
\mathcal{K}=\{K(z,q) \in \mathcal{M}^9 \ | \
K(z) \not\equiv 0, \ K(z) \  \textrm{satisfies \ the  \ reflection \ 
equation}  \ \eqref{re}  \}, 
\end{align}
where
\begin{align}
\mathcal{M}=\{ f(z,q) \ | \ f(z,q) \ \textrm{is meromorphic in}
\ z,q  \}.
\end{align}
We regard two solutions to be equivalent if they coincide up to 
an overall factor. 
We found two solution spaces 
$\mathcal{A}_{\mathrm{I}}$ and $\mathcal{A}_{\mathrm{II}}$
which can be parametrized by the following algebraic varieties,
$\mathcal{U}_{\mathrm{I}}$ and $\mathcal{U}_{\mathrm{II}}$.
\begin{definition}
Let $\mathcal{U}_{\mathrm{I}}, \mathcal{U}_{\mathrm{II}}$
be the following algebraic varieties.
\begin{align}
\mathcal{U}_{\mathrm{I}}&=
\mathbb{P}^1(\mathbb{C}) \times
\mathbb{P}^1(\mathbb{C}) \times
\mathbb{P}^2(\mathbb{C}), \\
\mathcal{U}_{\mathrm{II}}&=
\mathbb{C} \times
\mathbb{P}^1(\mathbb{C}) \times
\mathbb{P}^2(\mathbb{C}).
\end{align}
 \hspace*{\fill} \sq
\label{algebraicvariety}
\end{definition}
$\mathcal{A}_{\mathrm{I}}$ and $\mathcal{A}_{\mathrm{II}}$
is the space of $3 \times 3$ matrices which is parametrized by 
$\mathcal{U}_{\mathrm{I}}$ and $\mathcal{U}_{\mathrm{II}}$
respectively as follows.
\begin{definition}
We define two spaces of $3 \times 3$ matrices 
$\mathcal{A}_{\mathrm{I}}$ and $\mathcal{A}_{\mathrm{II}}$
as follows.
\begin{eqnarray}
\mathcal{A}_{\mathrm{I}}=\left\{ K_{\mathrm{I}}(z , 
\mathcal{P}_{\mathrm{I}}) \ \Big| \
\begin{array}{c}
\mathcal{P}_{\mathrm{I}}=
(B_1, B_2) \times (D_1, D_2) \times (E_1, E_2, E_3)
\in \mathcal{U}_{\mathrm{I}}
\end{array}
\right\}, \nn
\end{eqnarray}
\begin{align}
K_{\mathrm{I}}(z, \mathcal{P}_{\mathrm{I}})=&
K_{\mathrm{I},0}(z,D_1,D_2,E_1)-z^6 T K_{\mathrm{I},0}(z^{-1},D_2,D_1,E_3)T \nn \\
&+K_{\mathrm{I},1}(z,B_1,B_2,D_1,E_2)-z^6 T K_{\mathrm{I},1}(z^{-1},B_2,B_1,D_2,E_2)T,
\end{align}
where $K_{\mathrm{I},0}(z,D_1,D_2,E_1)$ and $K_{\mathrm{I},1}(z,B_1,B_2,D_1,E_2)$ are
\begin{align}
K_{\mathrm{I},0}(z,D_1,D_2,E_1)
=&E_1^2 \left(
\begin{array}{ccc}
D_2^2z^2  & D_1 D_2(z^4-1) &D_1^2  z^2(z^4-1) \\
0 &D_2^2 z^2 & D_1 D_2(z^4-1) \\
0 & 0 & D_2^2  z^2
\end{array}
\right), \\
K_{\mathrm{I},1}(z,B_1,B_2,D_1,E_2)
=&-D_1 E_2 z^2 \left(
\begin{array}{ccc}
B_1 & 0 & 0 \\
0 & B_1 & B_2(1-z^4) \\
0 & 0 & B_1 z^4
\end{array}
\right). 
\end{align}
\begin{eqnarray}
\mathcal{A}_{\mathrm{II}}=\left\{ K_{\mathrm{II}}(z , 
\mathcal{P}_{\mathrm{II}}) \ \Big| \
\begin{array}{c}
\mathcal{P}_{\mathrm{II}}=
(b) \times (F_1, F_2) \times (G_1, G_2, G_3)
\in \mathcal{U}_{\mathrm{II}}
\end{array}
\right\}, \nn
\end{eqnarray}
\begin{align}
K_{\mathrm{II}}(z, \mathcal{P}_{\mathrm{II}})=&
bz^2 \mathrm{Id}+ K_{\mathrm{II},0}(z,F_1,G_1,G_2,G_3)
-z^4 T K_{\mathrm{II},0}(z^{-1},F_2,G_3,-G_2,G_1)T,
\end{align}
where $K_{\mathrm{II},0}(z,F_1,G_1,G_2,G_3)$ is
\begin{align}
K_{\mathrm{II},0}(z,F_1,G_1,G_2,G_3)
=&-F_1 \left(
\begin{array}{ccc}
G_3  & 0 & G_1(1-z^4) \\
0 &G_3 & G_2(1-z^4) \\
0 & 0 & G_3 z^4
\end{array}
\right).
\end{align}
\hspace*{\fill} \sq
\label{defspace}
\end{definition}
The solutions to the reflection equation of the 
$N=3$ Cremmer-Gervais $R$-matrix can be expressed using the
spaces of $3 \times 3$ matrices 
$\mathcal{A}_{\mathrm{I}}$ and $\mathcal{A}_{\mathrm{II}}$
defined in Def \ref{defspace}.
\begin{theorem}{(Main Result)}
The solution space $\mathcal{K}$ of the 
$N=3$ Cremmer-Gervais $R$-matrix \eqref{cremmer-gervais}
is the union of $\mathcal{A}_{\mathrm{I}}$ and $\mathcal{A}_{\mathrm{II}}$,
and does not depend on the
parameters of the 
$R$-matrix, $q$.
\begin{equation}
\mathcal{K}=\mathcal{A}_{\mathrm{I}} \cup \mathcal{A}_{\mathrm{II}}. \nn
\end{equation}
\hspace*{\fill}    \sq
\label{maintheorem}
\end{theorem}
The following Proposition can be checked by a direct calculation.
\begin{proposition}
 $K_{\mathrm{I}}(z, \mathcal{P}_{\mathrm{I}})
 \in \mathcal{A}_{\mathrm{I}}$ and $
K_{\mathrm{II}}(z, \mathcal{P}_{\mathrm{II}})  \in \mathcal{A}_{\mathrm{II}}$
satisfy unitarity.
\begin{eqnarray}
K_{i}(z, \mathcal{P}_{i})K_{i}(z^{-1}, \mathcal{P}_{i})=\rho_{i}(z)
 \mathrm{Id}, \ \ \ i=\mathrm{I}, \mathrm{II},
\end{eqnarray}
where
\begin{align}
\rho_{\mathrm{I}}(z)=&  D_1^2(B_1^2+D_1^2)+D_2^2(B_2^2+D_2^2)+
(B_1 D_1^3+B_2 D_2^3-B_1 B_2 D_1 D_2)(z^2+z^{-2}) \nn \\
&-D_1 D_2(B_1 D_2+B_2 D_1)(z^4+z^{-4})
-D_1^2 D_2^2(z^6+z^{-6}),  \nn \\
\rho_{\mathrm{II}}(z)=&  b^2+F_2^2 G_1^2+F_1^2 G_3^2+b(F_2 G_1-F_1 G_3)(z^2+z^{-2})
+F_1 F_2 G_1 G_3(z^4+z^{-4}).  \nn
\end{align}  \hspace*{\fill}   \sq
\end{proposition}
We can also see that the solutions have the following 
transformation properties.
\begin{lemma}
We define the action $\mathrm{ad}$ as
$\mathrm{ad} \, X(Y)=XYX^{-1}$ for two $3\times3$ matrices $X$ and $Y$.
$K_{\mathrm{I}}(z, \mathcal{P}_{\mathrm{I}}) \in \mathcal{A}_{\mathrm{I}}$ and
$K_{\mathrm{II}}(z, \mathcal{P}_{\mathrm{II}}) \in \mathcal{A}_{\mathrm{II}}$
transform with respect to
the action of $G$ and $T$ \eqref{GT} as,
\begin{align}
\mathrm{ad} \, G(K_{i}(z, \mathcal{P}_{i}))&
=K_{i}(z,G \mathcal{P}_{i}),  \ \ \ i=\mathrm{I}, \mathrm{II},  \\
\mathrm{ad} \, T(K_{\mathrm{I}}(z, \mathcal{P}_{\mathrm{I}}))
&=z^{6} K_{\mathrm{I}}(z^{-1},T \mathcal{P}_{\mathrm{I}}), \\
\mathrm{ad} \, T(K_{\mathrm{II}}(z ,\mathcal{P}_{\mathrm{II}}))
&=z^{4} K_{\mathrm{II}}(z^{-1},T \mathcal{P}_{\mathrm{II}}),
\end{align}
where the action of
$G$ and $T$ on $\mathcal{P}_{\mathrm{I}}, \mathcal{P}_{\mathrm{II}}$ is defined by
\begin{align}
&G \mathcal{P}_{\mathrm{I}} 
=(B_1, \omega^2 B_2) \times (\omega^2 D_1, D_2) \times 
(E_1, \omega E_2, \omega E_3), \nn \\
&T \mathcal{P}_{\mathrm{I}} 
=-(B_2, B_1) \times (D_2, D_1) \times (E_3, E_2, E_1), \nn \\
&G \mathcal{P}_{\mathrm{II}} 
=(b) \times (\omega F_1, F_2) \times 
(G_1, \omega G_2, \omega^2 G_3), \nn \\
&T \mathcal{P}_{\mathrm{II}} 
=(b) \times (F_2, F_1) \times (-G_3, G_2, -G_1). \nn
\end{align}    \hspace*{\fill}    \sq
\end{lemma}
$G$ and $T$ act on the algebraic varieties
$\mathcal{U}_{\textrm{I}}$ and $ \mathcal{U}_{\textrm{II}}$,
and satisfy $G^3=\textrm{Id}$, $T^2=\textrm{Id}$ and  $(GT)^2=\textrm{Id}$.
\\
The common part of $\mathcal{A}_{\mathrm{I}} $ and $\mathcal{A}_{\mathrm{II}}$ is
\begin{lemma}
$\mathcal{A}_{\mathrm{I}} \cap \mathcal{A}_{\mathrm{II}}$, 
which is the common part of
$\mathcal{A}_{\mathrm{I}}$ and $ \mathcal{A}_{\mathrm{II}}$,
is
\begin{eqnarray}
&& \mathcal{A}_{\mathrm{I}} \cap \mathcal{A}_{\mathrm{II}}
= \mathcal{C} \cup \mathrm{ad} \, T 
(\mathcal{C}), \ \ \  \mathcal{C} \cap \mathrm{ad} \, T 
(\mathcal{C})= \{\mathrm{Id} \},
\\
&& \mathcal{C}=\left\{\left( 
\begin{array}{ccc}
c_{3}+c_{4}z^2 & 0 & 0 \\
c_{2}(z^4-1) & c_{4}z^2+c_{3}z^4 & 0 \\
c_{1}(z^4-1) & 0 & c_{4}z^2+c_{3}z^4
\end{array}
\right)  \right\},
\end{eqnarray}
where $(c_{1}, c_{2}) \times( c_{3}, c_{4})
\in \mathbb{C}^2 \times \mathbb{P}^1(\mathbb{C})$.
  \hspace*{\fill}   \sq  
\end{lemma}
\subsection{Properties of the Reflection  equation of the
$N=3$ Cremmer-Gervais $R$-matrix}
Utilizing the symmetries of the Cremmer-Gervais
$R$-matrix \eqref{conservation} and \eqref{t-invariance},
one can show that the solutions to the Reflection equation
has the following properties.
\begin{proposition}
$K(z,q)$ has the following properties.
\begin{align}
K(z,q) \in \mathcal{K}& \Rightarrow \mathrm{ad} \, G(K(z,q)) \in \mathcal{K},
\label{Kcon} \\
K(z,q) \in \mathcal{K}& \Rightarrow 
\mathrm{ad} \, T(K(z^{-1},q^{-1})) \in \mathcal{K}. \label{KT-inv1}
\end{align} \label{Kprop}  \hspace{\fill}   \sq
\end{proposition}
[ \textit{Proof of Prop \ref{Kprop}} ] \\
\eqref{Kcon} and \eqref{KT-inv1} follow from the symmetries of
$R^{CG}(z,q)$, \eqref{conservation} and \eqref{t-invariance},
respectively.
Since they can be proved similarly, we show \eqref{KT-inv1}.
Multiplying $T \otimes T$ from the left and right on both
sides of the reflection equation
\begin{align}
&R_{12}(z_{1}/z_{2},q)K_{1}(z_{1},q)R_{21}(z_{1} z_{2},q)K_{2}(z_{2},q) \nn \\
&=K_{2}(z_{2},q)R_{12}(z_{1} z_{2},q)K_{1}(z_{1},q)R_{21}(z_{1}/z_{2},q),
\label{refdef0}
\end{align}
and using \eqref{t-invariance}, one gets
\begin{align}
&R_{12}(z_{2}/z_{1},q^{-1})(T_{1}K_{1}(z_{1},q)T_{1})
R_{21}(z_{1}^{-1} z_{2}^{-1},q^{-1})(T_{2}K_{2}(z_{2},q)T_{2}) \nn \\
&=(T_{2}K_{2}(z_{2},q)T_{2})R_{12}(z_{1}^{-1} z_{2}^{-1},q^{-1})
(T_{1}K_{1}(z_{1},q)T_{1})R_{21}(z_{2}/z_{1},q^{-1}).
\label{refdef1}
\end{align}
Changing $z_{i} \to z_{i}^{-1}, q \to q^{-1}$ in
\eqref{refdef1}, we have
\begin{align}
&R_{12}(z_{1}/z_{2},q)(T_{1}K_{1}(z_{1}^{-1},q^{-1})T_{1})
R_{21}(z_{1} z_{2},q)(T_{2}K_{2}(z_{2}^{-1},q^{-1})T_{2}) \nn \\
&=(T_{2}K_{2}(z_{2}^{-1},q^{-1})T_{2})R_{12}(z_{1} z_{2},q)
(T_{1}K_{1}(z_{1}^{-1},q^{-1})T_{1})R_{21}(z_{1}/z_{2},q), 
\label{refdef2}
\end{align}
which means that \eqref{KT-inv1} holds.  \hspace*{\fill}  \wsq  \\
Let us investigate in more detail
the general properties of the reflection equation which comes from 
the symmetries of the Cremmer-Gervais $R$-matrix \eqref{t-invariance}.
From now on, we prepare some notations.
\begin{definition}  \hspace*{\fill} \\
$(1)$
We express the matrix elements of
$K(z) \in \mathrm{End}(\mathbb{C}^3)$
using $c_{j}^i(z,q)$ as
\begin{eqnarray}
K(z)=\left( 
\begin{array}{ccc}
c_{0}^{0}(z,q) & c_{1}^{0}(z,q) & c_{2}^{0}(z,q) \\
c_{0}^{1}(z,q) & c_{1}^{1}(z,q) & c_{2}^{1}(z,q) \\
c_{0}^{2}(z,q) & c_{1}^{2}(z,q) & c_{2}^{2}(z,q)
\end{array}
\right). \label{kexpress}
\end{eqnarray}
\\
$(2)$
We express the matrix elements of the matrix,
which is the left hand side of the reflection equation
subtracted by the right hand side, as
\begin{align}
(i_{1} i_{2}|j_{1} j_{2}):=&{[}  
R_{12}(z_{1}/z_{2},q)K_{1}(z_{1},q)R_{21}(z_{1} z_{2},q)K_{2}(z_{2},q)
{]}_{j_{1} j_{2}}^{i_{1} i_{2}} \nn \\
&-{[}
K_{2}(z_{2},q)R_{12}(z_{1} z_{2},q)
K_{1}(z_{1},q)R_{21}(z_{1}/z_{2},q) {]}_{j_{1} j_{2}}^{i_{1} i_{2}},
\end{align}
for $i_{1}, i_{2}, j_{1}, j_{2}=0,1,2$.  \\
$(3)$
 $T$ was defined as the matrix which acts on the vector space
$\mathbb{C}^3$, i.e., $T e_{j}=e_{2-j}$ $\mathrm{\eqref{GT}}$.
We also define the action of $T$ on the index $0,1,2$ of the
orthonormal basis $\{ e_{0}, e_{1}, e_{2} \}$ of 
$\mathbb{C}^3$ as
\begin{eqnarray}
T(j):=2-j,
\end{eqnarray}
for $j=0,1,2$. \hspace*{\fill}    \sq  \label{firstdef}
\end{definition}
Using the notations defined in
Def \ref{firstdef}, one has
\begin{eqnarray}
Te_j=e_{T(j)}, \ \ \
(\textrm{ad} \, T(K(z)))_{j}^i=K(z)_{T(j)}^{T(i)}, \ \ \
(\textrm{ad} \, T(R(z)))_{kl}^{ij}=R(z)_{T(k)T(l)}^{T(i)T(j)}.  \nn  
\end{eqnarray}
\begin{definition}
We define the space of meromorphic functions of
$z, q$ as $\mathcal{M}$.
\begin{eqnarray}
\mathcal{M}:=\left\{ f(z,q) \ \Big| \ 
f(z,q) \mathrm{\ is \
meromorphic \ in} \ z,q \right\}. \nn
\end{eqnarray}
$\mathcal{N}$ is defined as the 
space of homogeneous polynomials
of degree 1 with respect to $c_{j}^i(z_{1},q), i,j=0,1,2$, 
with respect to $c_{j}^i(z_{2},q), i,j=0,1,2$, and the coefficients 
belong to $\mathcal{M}$.
\begin{eqnarray}
\mathcal{N}=\left\{
g(c(z_{1}), c(z_{2}) |
z_{1}, z_{2}, q) \ \Big| \
\begin{array}{l}
g \ \mathrm{is \ a \ homogeneous \ polynomial \ of \ degree \ 1
\ with \ respect \ to }  \\
c_{j}^i(z_{1},q), i,j=0,1,2, \ \mathrm{with \ respect \ to \ }
c_{j}^i(z_{2},q), i,j=0,1,2 \\
\mathrm{and \ the \ coefficients \ belong \ to \ }
\mathcal{M}
\end{array} 
\right\}. \nn 
\end{eqnarray} \hspace*{\fill} \sq
\end{definition}
Expressing the matrix elements of
$K(z,q)$ as $c_{j}^i(z,q)$,
every element of the reflection equation
$(i_{1} i_{2}|j_{1} j_{2})$
belongs to $\mathcal{N}$, and can be expressed as
\begin{eqnarray}
\sum_{k,l,m,n=0}^{2}f_{kl, mn}(z_{1},z_{2},q)
c_{l}^{k}(z_{1},q)c_{n}^{m}(z_{2},q), \nn
\end{eqnarray}
where $f_{kl, mn}(z_{1},z_{2},q) \in \mathcal{M}$.
Comparing the reflection equation
\eqref{refdef0} with \eqref{refdef1},
we find the following holds.
\begin{lemma}
If an element of the reflection equation
$(i_{1} i_{2}|j_{1} j_{2})$
can be expressed as
\begin{eqnarray}
(i_1 i_2|j_1 j_2)=\sum_{k,l,m,n=0}^{2}f_{kl, mn}(z_{1},z_{2},q)
c_{l}^{k}(z_{1},q)c_{n}^{m}(z_{2},q),  \nn  \label{ex1}
\end{eqnarray}
where $c_{j}^{i}(z,q)$ are the matrix elements of $K(z,q)$
and $f_{kl, mn}(z_{1},z_{2},q) \in \mathcal{M}$, \\
$(T(i_{1}) T(i_{2})|T(j_{1})  T(j_{2}))$ can be expressed as
\begin{eqnarray}
\displaystyle  (T(i_{1}) T(i_{2})|T(j_{1})  T(j_{2}))=
\sum_{k,l,m,n=0}^{2}f_{kl, mn}(z_{1}^{-1},z_{2}^{-1},q^{-1})
c_{T(l)}^{T(k)}(z_{1},q)
c_{T(n)}^{T(m)}(z_{2},q). \nn  \label{ex2}
\end{eqnarray}
 \hspace*{\fill} \sq    \label{close}
\end{lemma}
We define the $T$-transformation of an element of
$\mathcal{N}$ as follows.
\begin{definition}
For an element of $\mathcal{N}$
\begin{eqnarray}
\sum_{k,l,m,n=0}^{2}f_{kl, mn}(z_{1},z_{2},q)
c_{l}^{k}(z_{1},q)c_{n}^{m}(z_{2},q),   \nn
\end{eqnarray}
where $f_{kl, mn}(z_{1},z_{2},q) \in \mathcal{M}$,
we define its $T$-transformation as
\begin{eqnarray}
&&T \left(
\sum_{k,l,m,n=0}^{2}f_{kl, mn}(z_{1},z_{2},q)
c_{l}^{k}(z_{1},q)c_{n}^{m}(z_{2},q)
\right)
 \nn \\
&& \ \ \ \ \ :=
\sum_{k,l,m,n=0}^{2}f_{kl, mn}(z_{1}^{-1},z_{2}^{-1},q^{-1})
c_{T(l)}^{T(k)}(z_{1},q)
c_{T(n)}^{T(m)}(z_{2},q).  \nn
\end{eqnarray}
 \hspace{\fill} \sq
\label{tdef}
\end{definition}
Defining the
$T$-transformation as above, Lemma \ref{close}
can be simply expressed as
\begin{corollary}
\begin{equation}
T(i_1 i_2|j_1 j_2)=(T(i_1)T(i_2)|T(j_1)T(j_2)).  \label{closecor}
\end{equation}
\hspace*{\fill} \sq
\end{corollary}
We also use the following Proposition
to prove Th \ref{maintheorem}, which is
about the subgroup of $\mathcal{N}$
invariant under the $T$-transformation.
\begin{proposition}
A subgroup $\tilde{\mathcal{N}}$ of $\mathcal{N}$
is called a {\it $T$-invariant subgroup}
if it is invariant under the $T$-transformation, i.e., 
if $T(g) \in \tilde{\mathcal{N}}$ holds for any
$g \in \tilde{\mathcal{N}}$,
If an element $g \in \mathcal{N}$
can be expressed as the linear combination
of elements of $\tilde{\mathcal{N}}$ with
$\mathcal{M}$-coefficients, i.e.,
can be expressed as
\begin{eqnarray}
g=
\sum_{\beta}h_{\beta}(z_{1}, z_{2}, q)
g_{\beta}(c(z_{1}), c(z_{2})
|z_{1}, z_{2}, q), \ \ 
g_{\beta} \in \tilde{\mathcal{N}}, \
h_{\beta} \in \mathcal{M}, 
 \label{beforeT2}
\end{eqnarray}
$T(g) \in \mathcal{N}$ can also be expressed as 
the linear combination of elements of $\tilde{\mathcal{N}}$
with $\mathcal{M}$-coefficients.
\begin{eqnarray}
T(g)=
\sum_{\beta}h_{\beta}(z_{1}^{-1}, z_{2}^{-1}, q^{-1})
T(g_{\beta}(c(z_{1}), c(z_{2})
|z_{1}, z_{2}, q)), \ \ 
T(g_{\beta})
 \in \tilde{\mathcal{N}}. 
\label{afterT}
\end{eqnarray} \label{KT-inv2} \hspace*{\fill}  \sq
\end{proposition}
We use \eqref{KT-inv1}, Lemma \ref{close}, 
the properties of the reflection equation coming 
from the symmetry of the Cremmer-Gervais $R$-matrix
\eqref{t-invariance},
and Prop \ref{KT-inv2},
the property which holds for the
$T$-invariant subgroup of
$\mathcal{N}$, to prove
Th \ref{maintheorem}.
\section{38 equations equivalent to the Reflection equation}
Solving the reflection equation is to solve 
81 equations $(i_{1} i_{2}|j_{1} j_{2})=0$,
$i_{1},i_{2},j_{1},j_{2}=0,1,2$ for $c_{j}^i(z,q), i,j=0,1,2$.
Directly solving the 81 equations is the most straightforward way,
which is a tiresome task. Instead,
carefully observing these 81 equations, one finds they
are equivalent to another set of 38 equations (Def \ref{38def}),
which is easier to handle.
This equivalence (Th \ref{38})
is shown in this section. We briefly outline the procedure.
First, we divide the 81 elements of the reflection equation
into several subgroups. Such is also done for the set of 38 equations.
Then we show 3 relations (Prop \ref{step0}, \ref{123123prime}, \ref{123ABC})
among the subgroups of the
reflection equation and the set of 38 equations.
Combining the relations and Prop \ref{KT-inv2}, we prove
the equivalence between the 81 equations $(i_{1} i_{2}|j_{1} j_{2})=0$,
$i_{1},i_{2},j_{1},j_{2}=0,1,2$ and another set of 38 equations.
\begin{theorem}
$\{(i_{1} i_{2}|j_{1} j_{2}) \ | \ i_{1},i_{2},j_{1},j_{2}=0,1,2 \}$,
which are the $81$ elements of the reflection equation of the
$N=3$ Cremmer-Gervais $R$-matrix $R^{CG}(z,q)$, 
are equivalent to $38$ equations which consist of
$20$ equations
$\{ A_j=0 \ | \  j=1, \cdots , 8 \}
 \cup \{ B_j=0 \ | \  j=1, \cdots , 7 \} \cup
\{ C_j=0 \ | \ j=1, \cdots , 5 \}$
defined in $\mathrm{Def} \ \ref{38def} $,
and their $T$-transformed $(\mathrm{Def} \ \ref{tdef})$
equations
$\{T\!A_j=0 \ | \  j=2,3,4,5,7,8 \} \cup 
\{T\!B_j=0 \ | \ j=1, \cdots , 7 \} \cup
 \{T\!C_j=0 \ | \ j=1, \cdots , 5 \}$.
\label{38}   \hspace*{\fill}   \sq
\end{theorem}
$A_{1}$ and $A_6$ are essentially self-dual 
with respect to  the
$T$-transformtion, i.e.,
$T\!A_{1}=-z_1^{-2} z_2^{-2} A_{1},
T\!A_{6}=z_{1}^{-4} z_{2}^{-4} A_{6}|_{z_{1} \leftrightarrow z_{2}}$.
\begin{definition}
We define 
$\{A_j,j=1, \cdots, 8 \}$,
$\{B_j,j=1, \cdots, 7 \}$,
$\{C_j, j=1, \cdots, 5 \}$
as the polynomials of $c_{j}^i(z,q)$,
the matrix elements of the
$K(z)$, $c_{j}^i(z,q)$, as follows.
For simplicity, we denote $c_{j}^i(z)=c_{j}^i(z,q)$.
\begin{align}
A_1:=& z_1^2 c_{1}^0(z_1) c_{1}^2(z_2)-
c_{1}^2(z_1) z_2^2 c_1^0(z_{2}), \nn  \\
A_2:=& c_{2}^0(z_1) z_{2}^2 c_{1}^0(z_2)-
z_{1}^2 c_1^0(z_1) c_2^0(z_2), \nn \\
A_3:=& c_{1}^2(z_1) c_{2}^0(z_2)
-c_2^0(z_1) c_1^2(z_2), \nn  \\
A_4:=& z_1^2 c_{1}^0(z_1) (c_{1}^0(z_2)-c_{2}^1(z_2))
-(c_{1}^0(z_1)-c_{2}^1(z_1)) z_2^2 c_1^0(z_2), \nn  \\
A_5:=& c_{1}^0(z_{1})c_{0}^1(z_{2})-c_{0}^1(z_{1})c_{1}^0(z_{2})
+c_{2}^0(z_{1})c_{0}^2(z_{2})-c_{0}^2(z_{1})c_{2}^0(z_{2}), \nn \\
A_6:=&(z_2^2-z_1^2)(z_1^2 z_2^2 c_1^0(z_1)(c_1^2(z_2)-c_0^1(z_2))-
c_1^2(z_2)(c_1^0(z_1)-c_2^1(z_1))) \nn \\
&+(z_1^4 c_0^0(z_1)-c_2^2(z_1))z_2^2(c_0^0(z_2)-c_2^2(z_2))
-z_1^2(c_0^0(z_1)-c_2^2(z_1))(z_2^4 c_0^0(z_2)-c_2^2(z_2)),
\nn \\
A_7:=&  c_{2}^1(z_{1})c_{1}^0(z_{2})-c_{1}^0(z_{1})c_{2}^1(z_{2})
+c_{2}^0(z_{1})(c_{0}^0(z_{2})-c_{2}^2(z_{2}))
-(c_{0}^0(z_{1})-c_{2}^2(z_{1}))c_{2}^0(z_{2}), \nn  \\
A_8:=&z_1^4 c_{1}^0(z_{1}) z_{2}^2 c_{2}^1(z_{2})
-z_{1}^2 c_{2}^1(z_{1}) z_2^4c_{1}^0(z_{2})
\nn \\
&+z_{2}^2 c_{2}^0(z_{2}) (z_1^4 c_{0}^0(z_{1})-c_{2}^2(z_{1}))
-(z_2^4 c_{0}^0(z_{2})-c_{2}^2(z_{2})) z_{1}^2 c_{2}^0(z_{1}), \nn \\
B_1:=& c_{1}^0(z_1) (c_{1}^1(z_2)-c_{0}^0(z_2))
-(c_{1}^1(z_1)-c_{0}^0(z_1)) c_1^0(z_2), \nn \\
B_2:=& c_{2}^0(z_1) (c_{1}^0(z_2)-c_{2}^1(z_2))
-(c_{1}^0(z_1)-c_{2}^1(z_1))c_2^0(z_2), \nn \\
B_3:=& c_{2}^0(z_1) (c_{1}^1(z_2)-c_{2}^2(z_2))
-(c_{1}^1(z_1)-c_{2}^2(z_1)) c_2^0(z_2), \nn \\
B_4:=& c_{1}^0(z_1) (c_{1}^2(z_2)-c_{0}^1(z_2))
-(c_{0}^1(z_1)-c_{1}^2(z_1)) c_1^0(z_2), \nn \\
B_5:=& z_{1}^2 c_{2}^0(z_{1})z_2^4 c_{0}^2(z_{2})
-z_1^4 c_{0}^2(z_{1}) z_{2}^2 c_{2}^0(z_{2})
+z_1^4 c_1^0(z_1) z_2^2 c_0^1(z_2)-z_1^2 c_0^1(z_1) z_2^4 c_1^0(z_2) \nn \\
&+(z_1^4 c_{1}^1(z_{1})- c_{2}^2(z_{1}))z_{2}^2(c_{1}^1(z_{2})-c_{2}^2(z_{2}))
-z_{1}^2(c_{1}^1(z_{1})-c_{2}^2(z_{1}))(z_2^4 c_{1}^1(z_{2})-c_{2}^2(z_{2})), \nn \\
B_6:=& c_2^0(z_1)c_0^1(z_2)-c_0^1(z_1) c_2^0(z_2)
+c_{2}^1(z_{1})(c_{1}^1(z_{2})-c_{2}^2(z_{2}))
-(c_{1}^1(z_{1})-c_{2}^2(z_{1})) c_{2}^1(z_{2}), \nn \\
B_7:=&z_{1}^2 c_{2}^1(z_{1})(z_2^4 c_{1}^1(z_{2})- c_{2}^2(z_{2}))
-(z_1^4 c_{1}^1(z_{1})-c_{2}^2(z_{1}))z_{2}^2 c_{2}^1(z_{2}) \nn \\
&+z_{1}^2 c_{2}^0(z_{1})(z_2^4 c_{0}^1(z_{2})-(z_{2}^4+1)c_{1}^2(z_{2}))
-(z_1^4 c_{0}^1(z_{1})-(z_{1}^4+1)c_{1}^2(z_{1})) z_{2}^2 c_{2}^0(z_{2}), \nn  
\end{align}
\begin{align}
C_1:=&  z_{1}^2 c_{1}^2(z_{1}) c_{2}^0(z_{2})
-c_{2}^0(z_{1}) z_{2}^2 c_{1}^2(z_{2}) \nn \\
&+z_{1}^2 c_{1}^0(z_{1}) (c_{1}^1(z_{2})-z_2^4 c_{0}^0(z_{2}))
-(c_{1}^1(z_{1})-z_1^4 c_{0}^0(z_{1})) z_{2}^2 c_{1}^0(z_{2}), \nn \\
C_2:=&z_1^2  c_{1}^0(z_1) (c_{1}^1(z_2)-c_{2}^2(z_2))
-(c_{1}^1(z_1)-c_{2}^2(z_1))z_2^2 c_1^0(z_2), \nn \\
C_3:=& c_{2}^0(z_{1}) z_2^2(c_{1}^2(z_{2})-c_{0}^1(z_{2}))
-z_1^2 (c_{1}^2(z_{1})-c_{0}^1(z_{1})) c_{2}^0(z_{2}) \nn \\
&+z_1^2(c_{1}^1(z_{1})-c_{0}^0(z_{1}))(c_{2}^1(z_{2})-c_{1}^0(z_{2}))
-(c_{2}^1(z_{1})-c_{1}^0(z_{1}))z_2^2 (c_{1}^1(z_{2})-c_{0}^0(z_{2})), \nn \\
C_4:=&  
c_{2}^0(z_{1})c_{0}^1(z_{2})-c_{0}^1(z_{1})c_{2}^0(z_{2}) 
+c_{1}^0(z_{1})(c_{1}^1(z_{2})-c_{2}^2(z_{2}))
-(c_{1}^1(z_{1})-c_{2}^2(z_{1}))c_{1}^0(z_{2}), \nn  \\
C_5:=&  z_1^2 z_2^4(c_0^1(z_1)c_2^1(z_2)-c_2^1(z_1)c_0^1(z_2)
+c_1^1(z_1)c_2^2(z_2)-c_2^2(z_1)c_1^1(z_2))
\nn \\
&+z_1^2 z_2^2(z_2^2-z_1^2)(c_1^0(z_1)c_0^1(z_2)+c_0^0(z_1)(c_1^1(z_2)-c_2^2(z_2)))
 \nn \\
&+c_2^2(z_1)z_2^2(c_1^1(z_2)-c_2^2(z_2))-z_1^2(c_1^1(z_1)-c_2^2(z_1))c_2^2(z_2). \nn 
\end{align}  \label{38def}  \hspace*{\fill}  \sq
\end{definition}
Since the coefficents of all the 38 equations
$\{ A_j=0 \ | \  j=1, \cdots ,  8\}
 \cup \{ B_j=0 \ | \  j=1, \cdots , 7 \} \cup
\{ C_j=0 \ | \ j=1, \cdots , 5 \}$,
$\{T\!A_j=0 \ | \  j=2, \cdots ,5,7, 8 \} \cup 
\{T\!B_j=0 \ | \ j=1, \cdots ,7 \} \cup
 \{T\!C_j=0 \ | \ j=1, \cdots , 5 \} \cup$
do not depend on the parameter of the
Cremmer-Gervais $R$-matrix, $q$,
$c_{j}^i(z,q)$ do not depend on  $q$.
Thus, we notice the following from $\mathrm{Th} \ \ref{38}$.
\begin{corollary}
The solution to the reflection equation of the
$N=3$ Cremmer-Gervais $R$-matrix 
does not depend on $q$.
 \label{38close}
\hspace*{\fill} \sq
\end{corollary}
Let us make some definitions
and prepare propositions to prove Th \ref{38}.
We first define the groups of polynomials.
\begin{definition}
We define
${\bf A}$, ${\bf B}$ and ${\bf C}$
as the following groups of polynomials.
\begin{align}
&{\bf A}:=\{A_{j} \ | \ j=1, \cdots , 8 \}, \nn \\
&{\bf B}:= \{B_{j} \ | \ j=1, \cdots , 7 \}, \nn \\
&{\bf C}:= \{C_{j} \ | \ j=1, \cdots , 5 \}. \nn  
\end{align}
For a  group of polynomials $J$,
let $T\! J$ be groups consisting of
polynomials which are the $T$-transformed polynomials
belonging to $J$.
\begin{align}
T\!J=\{ T\!X \ | \ X \in J \}.
\end{align}
For a group of polynomials $J$,
we also denote the groups of equations 
$\{X=0 \ | \  X \in J \}$ by $J$
as long as it does not make a confusion.
\hspace*{\fill} \sq
\end{definition}
\begin{definition}
We define 
${\bf 0}$,
${\bf 1}$,
${\bf 1^{\prime}}$,
${\bf 2}$,
${\bf 2^{\prime}}$,
${\bf 3}$ and
${\bf 3^{\prime}}$ as the following groups
of elements of the reflection equation.
\begin{align}
& {\bf 0}:=
\{(00|22), (00|11) \}, \nn \\
& {\bf 1}:=
\{(02|11), (00|21), (02|12), (01|21), 
(00|00), (02|20), (02|22), (20|22) \},  \nn \\
&{\bf 1^{\prime}}:=\{
(00|12),(10|12),(00|20),(00|02)
\}, \nn \\
&
{\bf 2}:=
\{
(20|21), (01|22), (01|12), (10|10),
(12|21), (21|22), (12|22) \}, \nn \\
&{\bf 2^{\prime}}:=\{
(10|22), (20|20), (21|12), (11|11), (21|21), (11|22), (10|21) \},  \nn \\
&{\bf 3}:=\{ 
(00|10), (20|12), (01|20), (01|02), (11|02) \}, \nn \\
&{\bf 3^{\prime}}:= \{
(01|11),(00|01),(10|11),(02|21),(10|02), (10|20), (11|12), (11|21) \}. \nn
\end{align}
For 
$J={\bf 0, 1, 2, 3, 1^{\prime}, 2^{\prime}, 3^{\prime}}$,
let $T\! J$ be groups of the elements of the reflection equation
as follows.
\begin{align}
T\!J=\{ (T(i_1) T(i_2)|T(j_1) T(j_2)) \ | \ (i_1 i_2|j_1 j_2) \in J \}.
\end{align}
\hspace*{\fill} \sq
\end{definition}
\begin{table}[htbp]
\begin{center}
\begin{eqnarray}
\left(
\begin{array}{ccccccccc}
{\bf 1} &{\bf 3^{\prime} }&{\bf 1^{\prime} }&{\bf 3 }&{\bf 0 }&{\bf 1^{\prime} }
&{\bf 1^{\prime} }&{\bf 1 }&{\bf 0} \\
T{\bf 2} &T{\bf 2^{\prime}} &{\bf 3} &T{\bf 2^{\prime} }&{\bf 3^{\prime}}&{\bf 2 }
&{\bf 3 }&{\bf 1 }&{\bf 2} \\
T{\bf 1} &T{\bf 2 }&T{\bf 2^{\prime}} &T{\bf 3 }&{\bf 1 }&{\bf 1} &{\bf 1} 
&{\bf 3^{\prime} }&{\bf 1 }\\
T{\bf 2 }&T{\bf 2 }&{\bf 3^{\prime} }&{\bf 2 }&{\bf 3^{\prime} }
&{\bf 1^{\prime} }&{\bf 3^{\prime} } 
&{\bf 2^{\prime} }&{\bf 2^{\prime} }\\
T{\bf 2^{\prime} }&T{\bf 3^{\prime} }&{\bf 3 }&T{\bf 3^{\prime} }&{\bf 2^{\prime} }
&{\bf 3^{\prime} }&T{\bf 3 }
&{\bf 3^{\prime} }&{\bf 2^{\prime} } \\
T{\bf 2^{\prime}} &T{\bf 2^{\prime} }&T{\bf 3^{\prime} }&T{\bf 1^{\prime} }
&T{\bf 3^{\prime} } 
&T{\bf 2 }&T{\bf 3^{\prime} }&{\bf 2} &{\bf 2 }\\
T{\bf 1} &T{\bf 3^{\prime} }&T{\bf 1 }&T{\bf 1 }&T{\bf 1 }&{\bf 3 }
&{\bf 2^{\prime} }&{\bf 2 }&{\bf 1 } \\
T{\bf 2 }&T{\bf 1 }&T{\bf 3 }&T{\bf 2 }&T{\bf 3^{\prime} }&{\bf 2^{\prime} }
&T{\bf 3 }&{\bf 2^{\prime} }&{\bf 2} \\
{\bf 0} &T{\bf 1 }&T{\bf 1^{\prime} }&T{\bf 1^{\prime} }&{\bf 0 }&T{\bf 3 }
&T{\bf 1^{\prime} }&T{\bf 3^{\prime}}&T{\bf 1} 
\end{array}
\right) \nn
\end{eqnarray}
\end{center}
\caption{The elements of the reflection equation
$(i_{1} i_{2}|j_{1} j_{2})$ and their associated groups.
The matrix element $(3i_1+i_2+1, 3j_1+j_2+1)$ is 
assigned to $(i_{1} i_{2}|j_{1} j_{2})$.}
\end{table}
For these groups of polynomials, we note the following.
\begin{lemma} 
$(\mathrm{i})$ \ $J \cup T\!J$
is invariant under the $T$-transformation for each 
$J={\bf A}, {\bf B}, {\bf C}$. \\
$(\mathrm{ii})$ \ 
$J \cup T\!J$
is invariant under the $T$-transformation for each
$J={\bf 0, 1, 2, 3, 1^{\prime}, 2^{\prime}, 3^{\prime}}$.
\label{01234inv}
\hspace*{\fill} \sq
\end{lemma}
$(\mathrm{i})$ is obvious from the definition.
We also note $(\mathrm{ii})$
from the fact that $T(i_1 i_2|j_1 j_2)$, which is the
$T$-transformation of $(i_1 i_2|j_1 j_2)$, is $(T(i_1)T( i_2)|T(j_1)T( j_2))$
\eqref{closecor}. \\
We introduce the following notations for simplicity.
\begin{definition}
Let {\bf Reflection}, {\bf Reduced} be the following groups of polynomials.
\begin{align}
&{\bf Reflection}=\{(i_1 i_2|j_1 j_2) \ | \ i_1, i_2, j_1, j_2=0,1,2  \}, \nn \\
&{\bf Reduced}= {\bf A \cup B  \cup C} \cup T\!{\bf A} \cup T\! {\bf B} \cup T\! {\bf C}.
\nn
\end{align}
\hspace*{\fill} \sq
\end{definition}
By definition, we obviously have
\begin{align}
{\bf Reflection}=\bigcup_{J=0,1,2,3,1^{\prime}, 2^{\prime}, 3^{\prime}} \{ J \cup T\!J \}. \nn
\end{align}
For two groups of polynomials $P$ and $Q$,
let us denote $P \Rightarrow Q$
if all the polynomials in
 $Q$ can be expressed as linear combinations of
the polynomials in $P$
with $\mathcal{M}$-coefficients.
In order to prove Th \ref{38},
we prepare three Propositions about the relations between the
groups of polynomials which have been just defined.
The proof of these Propositions is given in Appendix A.
\begin{proposition}
The $4$ elements which belong to
${\bf 0} \cup T\!{\bf 0}$ are all $0$.
\hspace*{\fill}  \sq
\label{step0}
\end{proposition}
\begin{proposition}
\begin{eqnarray}
{\bf 1 \cup 2 \cup 3 \cup} T\!{\bf 2}
\Longrightarrow {\bf 1^{\prime} \cup 2^{\prime} \cup 3^{\prime}}. \nn
\end{eqnarray}
  \hspace*{\fill} \sq \label{123123prime}
\end{proposition}
\begin{proposition}
\begin{eqnarray}
{\bf 1 \cup 2 \cup 3}
\Longleftrightarrow {\bf A \cup B \cup C}. \nn
\end{eqnarray}
  \hspace*{\fill}   \sq \label{123ABC}
\end{proposition}
[ \textit{Proof of Th \ref{38}}  ] \\
Th \ref{38} means
\begin{eqnarray}
{\bf Reflection} \Longleftrightarrow {\bf Reduced}, \label{simplify}
\end{eqnarray}
which can be decomposed into
\begin{eqnarray}
&(\mathrm{i}) \ {\bf Reflection} \Longrightarrow {\bf Reduced}, \nn \\
&(\mathrm{ii}) \ {\bf Reduced} \Longrightarrow {\bf Reflection}. \nn
\end{eqnarray}
We can prove $(\mathrm{i})$ and $(\mathrm{ii})$ by utilizing
Prop \ref{123123prime}, \ref{123ABC} and \ref{KT-inv2}. \\
$(\mathrm{i})$
From Prop \ref{123ABC}, we have
\begin{eqnarray}
{\bf 1 \cup 2 \cup 3}
\Longrightarrow {\bf A \cup B \cup C}. \nn
\end{eqnarray}
Combining this with the obvious relation
\begin{eqnarray}
{\bf Reflection}
\Longrightarrow {\bf 1 \cup 2 \cup 3}, \nn
\end{eqnarray}
one has
\begin{eqnarray}
{\bf Reflection}
\Longrightarrow {\bf A \cup B \cup C}. \label{refabc}
\end{eqnarray}
Setting $\tilde{\mathcal{N}}={\bf Reflection}$ in
Prop \ref{KT-inv2}, \eqref{refabc} gives
\begin{eqnarray}
{\bf Reflection}
\Longrightarrow T\!{\bf A} \cup T\!{\bf B} \cup T\!{\bf C}. \label{reftabc}
\end{eqnarray}
Combining \eqref{refabc} and \eqref{reftabc} together, one has 
$(\mathrm{i})$. \\
$(\mathrm{ii})$ 
From Prop \ref{123ABC}, one has
\begin{eqnarray}
{\bf A \cup B \cup C}
\Longrightarrow {\bf 1 \cup 2 \cup 3}. \nn
\end{eqnarray}
Combining this with 
\begin{eqnarray}
{\bf Reduced}
\Longrightarrow {\bf A \cup B \cup C}, \nn
\end{eqnarray}
which is an obvious relation, we have
\begin{eqnarray}
{\bf Reduced}
\Longrightarrow {\bf 1 \cup 2 \cup 3}. \label{red123}
\end{eqnarray}
Setting $\tilde{\mathcal{N}}={\bf Reduced}$ in
Prop \ref{KT-inv2}, one gets
\begin{eqnarray}
{\bf Reduced}
\Longrightarrow T\!{\bf 1} \cup T\!{\bf 2} \cup T\!{\bf 3}, \label{redt123}
\end{eqnarray}
from \eqref{red123}.
Combining
\eqref{red123} and \eqref{redt123} gives
\begin{eqnarray}
{\bf Reduced}
\Longrightarrow {\bf 1 \cup 2 \cup 3}\cup
T\!{\bf 1} \cup T\!{\bf 2} \cup T\!{\bf 3}. \label{red123t123}
\end{eqnarray}
We combine 
Prop \ref{123123prime}, \eqref{red123t123} and
$
{\bf 1 \cup 2 \cup 3}\cup
T\!{\bf 1} \cup T\!{\bf 2} \cup T\!{\bf 3}
\Longrightarrow {\bf 1 \cup 2 \cup 3}\cup T\!{\bf 2}
$
to get
\begin{eqnarray}
{\bf Reduced}
\Longrightarrow {\bf 1^{\prime} \cup 2^{\prime} \cup 3^{\prime}}. \label{red123prime}
\end{eqnarray}
Setting $\tilde{\mathcal{N}}={\bf Reduced}$ in
Prop \ref{KT-inv2}, \eqref{red123prime} leads to
\begin{eqnarray}
{\bf Reduced}
\Longrightarrow 
T\!{\bf 1^{\prime}} \cup T\!{\bf 2^{\prime}} \cup T\!{\bf 3^{\prime}}.
\label{redt123prime}
\end{eqnarray}
From \eqref{red123prime} and \eqref{redt123prime}, we have
\begin{eqnarray}
{\bf Reduced}
\Longrightarrow {\bf 1^{\prime} \cup 2^{\prime} \cup 3^{\prime}}\cup
T\!{\bf 1^{\prime}} \cup T\!{\bf 2^{\prime}} \cup T\!{\bf 3^{\prime}}.
\label{red123t123prime}
\end{eqnarray}
Combining \eqref{red123t123} and \eqref{red123t123prime}, one gets 
$(\mathrm{ii})$.
\hspace*{\fill} \wsq \\
To determine the solution space,
we use the following 38 equations ${\bf Reduced^{\prime}}$
instead of ${\bf Reduced}$ since they are  easier to treat.
\begin{definition}
Let 
$A_5^{\prime}, A_6^{\prime}$ and  $C_5^{\prime}$ be the following polynomials.
\begin{align}
A_5^{\prime}:=&A_5+B_4, \nn \\
A_6^{\prime}:=&z_1^4 c_1^0(z_1) z_2^2 c_0^1(z_2)-z_1^2 c_0^1(z_1) z_2^4 c_1^0(z_2)
+c_1^2(z_1) z_2^2 c_2^1(z_2)-z_1^2 c_2^1(z_1) c_1^2(z_2), \nn \\
&+(z_1^4 c_0^0(z_1)-c_2^2(z_1))z_2^2(c_0^0(z_2)-c_2^2(z_2))
-z_1^2(c_0^0(z_1)-c_2^2(z_1))z_2^4 c_0^0(z_2)-c_2^2(z_2)), \nn \\
C_5^{\prime}:=&c_0^1(z_1)(c_1^0(z_2)-c_2^1(z_2))-(c_1^0(z_1)-c_2^1(z_1))c_0^1(z_2) \nn \\
&+(c_0^0(z_1)-c_1^1(z_1))(c_2^2(z_2)-c_1^1(z_2))
-(c_2^2(z_1)-c_1^1(z_1))(c_0^0(z_2)-c_1^1(z_2)). \nn
\end{align}
Let 
${\bf A^{\prime}, C^{\prime}}$ be the following groups of polynomials.
\begin{align}
{\bf A^{\prime}}&:=\{ A_j \ | \ j=1, \cdots,7,8 \} \cup \{A_5^{\prime},
A_6^{\prime} \}, \nn \\
{\bf C^{\prime}}&:=\{ C_j \ | \ j=1, \cdots , 4 \} \cup \{ C_5^{\prime} \}. \nn
\end{align}
Let 
${\bf Reduced^{\prime}}$ be a group consisting of the following 
$38$ polynomials.
\begin{align}
{\bf Reduced^{\prime}}= {\bf A^{\prime} \cup B  \cup C^{\prime}} \cup 
T\!{\bf A^{\prime}} \cup T\! {\bf B} \cup T\! {\bf C^{\prime}}. \nn
\end{align}
\hspace*{\fill} \sq
\end{definition}
$A_6^{\prime}$ is essentially self-dual
with respect to the $T$-transformation, i.e.,
$T\!A_6^{\prime}=z_1^{-4} z_2^{-4} A_6^{\prime}$.
\begin{proposition}
\begin{align}
{\bf Reduced} \Longleftrightarrow {\bf Reduced^{\prime}}.
\end{align}
\hspace*{\fill} \sq \label{redredprime}
\end{proposition}
{[} \textit{Proof of Prop \ref{redredprime}} {]} \\
This follows from
\begin{align}
A_5, B_4 &\Longleftrightarrow
A_5^{\prime}, B_4, \\
T\!A_5, T\!B_4 &\Longleftrightarrow
T\!A_5^{\prime}, T\!B_4, \\
A_1, B_4, T\!B_4, A_6 &\Longleftrightarrow
A_1, B_4, T\!B_4, A_6^{\prime}, \label{a6a6prime} \\
A_6, B_5, T\!B_5, C_5 &\Longleftrightarrow
A_6, B_5, T\!B_5, C_5^{\prime}, \\
T\!A_1, B_4, T\!B_4, T\!A_6 &\Longleftrightarrow
T\!A_1, B_4, T\!B_4, T\!A_6^{\prime},  \\
T\!A_6, B_5, T\!B_5, T\!C_5 &\Longleftrightarrow
T\!A_6, B_5, T\!B_5, T\!C_5^{\prime}.
\end{align}
For example, \eqref{a6a6prime} holds since $A_6^{\prime}$ can be expressed
using $A_1, B_4, T\!B_4$ and $A_6$ as
\begin{align}
A_6^{\prime}=A_6+(z_1^2 z_2^2-1)A_1-z_1^2 z_2^4 B_4-z_2^2 T\!B_4. \nn
\end{align}
\hspace*{\fill} \wsq \\
\section{Solving the Reflection equation}
In this section, we prove Th \ref{maintheorem}, i.e.,
determine the solution space $\mathcal{K}$
of the reflection equation by solving the 38 equations
${\bf Reduced^{\prime}}$.
The ouline is as follows.
We introduce 9 solution subspaces
$\mathcal{K}_j^i, i,j=0,1,2$, where
$\mathcal{K}_j^i(\subset \mathcal{M}^9 \setminus \{ \mathbf{0} \},
\mathcal{M}:$ 
space of meromorphic functions of $z,q$) is an open space
whose matrix element
$c_j^i(z)$ is not identically 0.
In order to determine the full solution space $\mathcal{K}$,
we construct the union of 9 open solution
subspaces $\mathcal{K}_j^i, i,j=0,1,2$,
which is a cover of $\mathcal{K}$.
As a result, the solution space is shown (Th \ref{subtheorem})
to be the union of two subspaces $\mathcal{B}_{\mathrm{I}}$ and
$\mathcal{B}_{\mathrm{II}}$ (Def \ref{spacedef3}),
which can be parametrized by
algebraic varieties $\mathcal{V}_{\mathrm{I}}$ and $\mathcal{V}_{\mathrm{II}}$
(Def \ref{spacedef2}). Analyzing these varieties,
we find (Th \ref{maintheorem}) the solution space is the union of two subspaces
$\mathcal{A}_{\mathrm{I}}$ and
$\mathcal{A}_{\mathrm{II}}$ (Def \ref{defspace}), each of which
can be parametrized by $\mathbb{P}^1(\mathbb{C}) \times
\mathbb{P}^1(\mathbb{C}) \times
\mathbb{P}^2(\mathbb{C})$
and
$
\mathbb{C} \times
\mathbb{P}^1(\mathbb{C}) \times
\mathbb{P}^2(\mathbb{C})$.
\begin{definition}
We define
$\mathcal{V}_{\mathrm{I}}$ and $\mathcal{V}_{\mathrm{II}}$ 
as the following algebraic variety in
$\mathbb{P}^9(\mathbb{C})$ and $\mathbb{P}^6(\mathbb{C})$,
respectively.
\begin{eqnarray}
\mathcal{V}_{\mathrm{I}}=\left\{
\left(
\begin{array}{ccccc}
a_{0} & a_{1} & a_{2} & a_{3} & a_{4} \\
\bar{a}_{0} & \bar{a}_{1} & \bar{a}_{2} & \bar{a}_{3} & \bar{a}_{4} 
\end{array}
\right) \in \mathbb{P}^9(\mathbb{C}) \ \Big| \
\begin{array}{l}
a_{j}, \ \bar{a}_{j}, \ j=0 \sim 4
 \ 
 \textrm{satisfy} \ 14 \ \textrm{relations in} \
\eqref{relation1}
 \\
(\mathrm{I}_j, \ j=1, \cdots , 8), \
(T\mathrm{I}_j, \  j=3, \cdots , 8 ) 
\end{array} 
\right\}, 
\label{space1}
\end{eqnarray}
\begin{eqnarray}
\begin{array}{cccccc}
 \mathrm{I}_1: \ a_{0} \bar{a}_{0}-a_{1} \bar{a}_{1}=0, & 
 \mathrm{I}_2: \ a_{2} \bar{a}_{2}-a_{3} \bar{a}_{3}=0, & 
 \mathrm{I}_3: \ a_{0}^2-a_{1} a_{4}=0,  \\
 T \mathrm{I}_3: \ \bar{a}_{0}^2-\bar{a}_{1} \bar{a}_{4}=0, &
 \mathrm{I}_4: \ a_{1} \bar{a}_{0}-a_{0} \bar{a}_{4}=0, &
 T \mathrm{I}_4: \ \bar{a}_{1} a_{0}-\bar{a}_{0} a_{4}=0, \\
 \mathrm{I}_5: \ a_{0} a_{3}-a_{1} \bar{a}_{2}=0, &
 T \mathrm{I}_5: \ \bar{a}_{0} \bar{a}_{3}-\bar{a}_{1} a_{2}=0, &
 \mathrm{I}_6: \ a_{0} a_{2}-\bar{a}_{3} a_{1}=0, \\
 T \mathrm{I}_6: \ \bar{a}_{0} \bar{a}_{2}-a_{3} \bar{a}_{1}=0, &  
 \mathrm{I}_7: \ a_{2} \bar{a}_{0}-\bar{a}_{3} \bar{a}_{4}=0, & 
 T \mathrm{I}_7: \ \bar{a}_{2} a_{0}-a_{3} a_{4}=0, \\
 \mathrm{I}_8:  \ a_{2} a_{4}-a_{0} \bar{a}_{3}=0, &
 T \mathrm{I}_8: \ \bar{a}_{2} \bar{a}_{4}-\bar{a}_{0} a_{3}=0.
\end{array} 
\label{relation1}
\end{eqnarray}
\begin{eqnarray}
\mathcal{V}_{\mathrm{II}}=\left\{
(b, b_{0}, b_{1}, b_{2}, b_{3}, b_{4}, b_{5})
\in \mathbb{P}^6 (\mathbb{C}) \ \Big| \
\begin{array}{l}
b, \ b_{j}, \ j=0 \sim 5
 \ \textrm{satisfy} \ 3 \ \textrm{relations in}  
\ \eqref{relation2}
 \\
\mathrm{II}_1, \mathrm{II}_2, \mathrm{II}_3 
\end{array}
\right\},
\label{space2}
\end{eqnarray}
\begin{eqnarray}
\begin{array}{cccccc}
\mathrm{II}_1: \ b_{0} b_{1}-b_{3} b_{4}=0, &
\mathrm{II}_2: \ b_{1} b_{2}-b_{4} b_{5}=0, &
\mathrm{II}_3: \ b_{2} b_{3}-b_{5} b_{0}=0.
\end{array}
\label{relation2}
\end{eqnarray} \hspace*{\fill} \sq
\label{spacedef2}
\end{definition}
\begin{definition}
Let 
$\mathcal{B}_{\mathrm{I}}$ and $\mathcal{B}_{\mathrm{II}}$
be the following spaces of $3 \times 3$ matrices.
\begin{eqnarray}
\mathcal{B}_{\mathrm{I}}=\left\{ K_{\mathrm{I}}(z , \mathcal{Q}_{\mathrm{I}})
\ \Big| \
\mathcal{Q}_{\mathrm{I}}=
\left(
\begin{array}{ccccc}
a_{0} & a_{1} & a_{2} & a_{3} & a_{4} \\
\bar{a}_{0} & \bar{a}_{1} & \bar{a}_{2} & \bar{a}_{3} & \bar{a}_{4} 
\end{array}
\right)
\in \mathcal{V}_{\mathrm{I}} \  \eqref{space1} 
\right\}, 
\end{eqnarray}
where \\
$K_{\mathrm{I}}(z,\mathcal{Q}_{\mathrm{I}}) $
\begin{eqnarray}
=\left(
\begin{array}{ccc}
\bar{a}_{3}+(a_{4}-a_{3})z^2 -\bar{a}_{4} z^4 & a_{0}(z^4-1) & a_{1}z^2(z^4-1)
\\
(\bar{a}_{2}+\bar{a}_{0}z^2)(z^4-1) &
(a_{4}-a_{3})z^2-(\bar{a}_{4}-\bar{a}_{3})z^4 & 
(a_{0}+a_{2}z^2)(z^4-1) \\
\bar{a}_{1}(z^4-1) & \bar{a}_{0}z^2(z^4-1) &
a_{4}z^2-(\bar{a}_{4}-\bar{a}_{3})z^4-a_{3}z^6 
\end{array}
\right). \label{first}
\end{eqnarray}
\begin{eqnarray}
\mathcal{B}_{\mathrm{II}}=\left\{ K_{\mathrm{II}}(z , \mathcal{Q}_{\mathrm{II}}) 
\ \Big| \
\mathcal{Q}_{\mathrm{II}}=(b, b_{0}, b_{1}, b_{2}, b_{3}, b_{4}, b_{5})
\in  \mathcal{V}_{\mathrm{II}} \
\eqref{space2}
\right\},
\end{eqnarray}
where
\begin{eqnarray}
 K_{\mathrm{II}}(z, \mathcal{Q}_{\mathrm{II}}) 
=\left(
\begin{array}{ccc}
b_{3}-b_{4}+b z^2 & 0 & b_{0}(z^4-1) \\
-b_{5}(z^4-1) & -b_{4}+b z^2 +b_{3} z^4 & b_{2}(z^4-1) \\
b_{1}(z^4-1) & 0 & b z^2+(b_{3}-b_{4})z^4
\end{array}
\right). \label{second}
\end{eqnarray}
\label{spacedef3}
\hspace*{\fill}    \sq
\end{definition}
In order to prove Th \ref{maintheorem},
we first prove the following Theorem.
\begin{theorem}
The solution space $\mathcal{K}$ of the 
$N=3$ Cremmer-Gervais $R$-matrix \eqref{cremmer-gervais}
is the union of $\mathcal{B}_{\mathrm{I}}$, $\mathcal{B}_{\mathrm{II}}$,
and does not depend on the parameter of the 
$R$-matrix, $q$.
\begin{equation}
\mathcal{K}=\mathcal{B}_{\mathrm{I}} \cup \mathcal{B}_{\mathrm{II}}. \nn
\end{equation}
\hspace*{\fill}    \sq
\label{subtheorem}
\end{theorem}
{[} \textit{Proof of Th \ref{subtheorem}} {]} \\
From
Th \ref{38} and Prop \ref{redredprime},
it is enough to solve the 38 equations ${\bf Reduced^{\prime}}$
to determine the solution space of the reflection equation.
One notices $K(z,q)=K(z)$ from
Cor \ref{38close}.
Since we are considering solutions which are not identically 0,
at least one of the matrix elements of the relection equation
is not identically 0. Then we have
\begin{equation}
\mathcal{K}= \bigcup_{i,j=0}^{2} \mathcal{K}_{j}^i, \label{setprop}
\end{equation}
where 
\begin{align}
\mathcal{K}_{j}^i&:=\mathcal{K} \cap U_{j}^i, \nn \\
U_{j}^i&:=\{ K(z)=(c_{l}^k(z))_{k,l=0,1,2} \in \mathcal{M}^9
\ | \ c_{j}^i(z) 
\not\equiv 0    \} \subset \mathcal{M}^9 \setminus \{ \mathbf{0} \}. \nn
\end{align}
Thus, if we determine
9 subspaces
$\mathcal{K}_{j}^i, i,j=0,1,2$,
one can get 
$\mathcal{K}$
as their union. \\
For a solution space $\mathcal{K}_{l}^k$,
we define its $T$-transformed space
$T(\mathcal{K}_{l}^k)$ as follows.
\begin{definition}
\begin{eqnarray}
T(\mathcal{K}_l^k)=\{ K^{\prime}(z)=TK(z^{-1})T \  | \  K(z) \in \mathcal{K}_{l}^k \}.
\nn
\end{eqnarray}
\hspace*{\fill} \sq
\end{definition}
From \eqref{KT-inv1}, it is easy to see
\begin{eqnarray}
T(\mathcal{K}_{l}^k)=\mathcal{K}_{T(l)}^{T(k)}. \label{solutionspaceT}
\end{eqnarray}
From \eqref{solutionspaceT},
the solution space $\mathcal{K}_{T(l)}^{T(k)}=\mathcal{K}_{2-l}^{2-k}$
can be obtained from $\mathcal{K}_{l}^k$.
Thus, in order to obtain $\mathcal{K}$,
it is enough to determine 6 spaces
\begin{equation}
\mathcal{K}_{0}^0, \mathcal{K}_{1}^0, \mathcal{K}_{2}^0, 
\mathcal{K}_{1}^1, \mathcal{K}_{2}^1, \mathcal{K}_{2}^2, \label{6eq}
\end{equation}
out of 9 spaces $\mathcal{K}_{j}^i,i,j=0,1,2$, and
$\mathcal{K}_1^2, \mathcal{K}_0^2$ and  $\mathcal{K}_0^1$
can be easily obtained from $\mathcal{K}_1^0, \mathcal{K}_2^0$ and $\mathcal{K}_2^1$
respectively.
Let us further consider solution spaces
$\mathcal{K}_0^0 \cup \mathcal{K}_1^1 \cup \mathcal{K}_2^2$
whose diagonal elements are not identically 0.
Since several equations in ${\bf Reduced^{\prime}}$
have terms
$c_{1}^1(z)-c_{0}^0(z),c_{1}^1(z)-z^4 c_{0}^0(z),c_{1}^1(z)-c_{2}^2(z)$,
it is easier to handle the solution spaces
in which these terms are not identically 0. 
Such solution spaces are equivalent to $\mathcal{K}_0^0 \cup \mathcal{K}_1^1
\cup \mathcal{K}_2^2$.
\begin{lemma}
\begin{eqnarray}
\mathcal{K}_0^0 \bigcup \mathcal{K}_1^1 \bigcup \mathcal{K}_2^2=
\mathcal{D}^0 \bigcup \mathcal{D}^1 \bigcup \mathcal{D}^2, \nn
\end{eqnarray}
where 
\begin{align}
\mathcal{D}^j&:=\mathcal{K} \cap V^j, j=0,1,2, \nn \\
V^0&:=\{ K(z)=(c_{l}^k(z))_{k,l=0,1,2} \in \mathcal{M}^9 \setminus \{ \mathbf{0} \}
\ | \ c_{1}^1(z)-c_{0}^0(z)
\not \equiv 0 \}, \nn  \\
V^1&:=\{ K(z)=(c_{l}^k(z))_{k,l=0,1,2} \in
\mathcal{M}^9 \setminus \{ \mathbf{0} \}
 \ | \ c_{1}^1(z)-z^4 c_{0}^0(z)
\not \equiv 0 \}, \nn \\
V^2&:=
\{ K(z)=(c_{l}^k(z))_{k,l=0,1,2} \in \mathcal{M}^9 \setminus \{ \mathbf{0} \}
\ | \ c_{1}^1(z)-c_{2}^2(z)
\not \equiv 0 \}. \nn
\end{align}
\hspace*{\fill} \sq \label{diagonaleq}
\end{lemma}
{[}\textit{Proof of Lemma \ref{diagonaleq}} {]}\\
This follows from 
\begin{eqnarray}
U_0^0 \bigcup U_1^1 \bigcup U_2^2=
V^0 \bigcup V^1 \bigcup V^2. \nn
\end{eqnarray}
\hspace*{\fill} \wsq \\
From 
\eqref{solutionspaceT} and Lemma \ref{diagonaleq}, one has
\begin{proposition}
\begin{align}
\mathcal{K}=&\mathcal{K}_{1}^0 \bigcup T(\mathcal{K}_{1}^0) \bigcup 
\mathcal{K}_{2}^0 \bigcup T(\mathcal{K}_{2}^{0})
 \bigcup \mathcal{K}_{2}^{1} \bigcup 
T(\mathcal{K}_{2}^{1}) \bigcup \mathcal{D}, \nn 
\end{align}
where $\mathcal{D}=\mathcal{D}^{0} \cup \mathcal{D}^{1} \cup \mathcal{D}^{2}$.
\hspace*{\fill} \sq
\end{proposition}
Thus, one can obtain the solution space $\mathcal{K}$ by determining
\begin{equation}
\mathcal{K}_{1}^0, \mathcal{K}_{2}^0, \mathcal{K}_{2}^1, \mathcal{D}, \nn
\end{equation}
instead of \eqref{6eq}.
Among
$\mathcal{K}_{1}^0, \mathcal{K}_{2}^0, \mathcal{K}_{2}^1, \mathcal{D}$,
we first determine $\mathcal{K}_{1}^0$ and $\mathcal{K}_{2}^0$.
The results, together with those for $\mathcal{K}_{1}^2$ and $\mathcal{K}_{0}^2$
can be stated as
\begin{proposition}
\begin{align}
& (\mathrm{i}) \ \mathcal{K}_{1}^0=\mathcal{B}_{\mathrm{I}}|_{a_0 \neq 0}. \nn \\
& (\mathrm{ii}) \ \mathcal{K}_{2}^0=\mathcal{B}_{\mathrm{I}}|_{a_1 \neq 0} \cup
\mathcal{B}_{\mathrm{II}}|_{b_0 \neq 0}. \nn \\
& (\mathrm{iii}) \ \mathcal{K}_{1}^2=\mathcal{B}_{\mathrm{I}}|_{\bar{a}_0 \neq 0}. \nn \\
& (\mathrm{iv}) \ \mathcal{K}_{0}^2=\mathcal{B}_{\mathrm{I}}|_{\bar{a}_1 \neq 0} \cup
\mathcal{B}_{\mathrm{II}}|_{b_1 \neq 0}. \nn
\end{align}
Here, for example, $\mathcal{B}_{\mathrm{I}}|_{a_0 \neq 0}$ is the 
subspace of $\mathcal{B}_{\mathrm{I}}$
which satisfies $a_0 \neq 0$.
\hspace*{\fill}    \sq
\label{det02}
\end{proposition}
Utilizing \eqref{solutionspaceT},
one can obtain $\mathcal{K}_1^2$ and $\mathcal{K}_0^2$ from
$\mathcal{K}_1^0$ and $\mathcal{K}_2^0$ respectively.
Thus, to prove Prop \ref{det02}, it is enough to compute 
$\mathcal{K}_{1}^0$ and $\mathcal{K}_{2}^0$.
Prop \ref{det02} (ii) is proved in Appendix B. \\
Next, we determine $\mathcal{K}_2^1$ and $\mathcal{K}_0^1$.
Among $\mathcal{K}_2^1$ and $\mathcal{K}_0^1$,
solutions which satisfy
$c_1^0(z) \not\equiv 0$ or $c_2^0(z) \not\equiv 0$ or 
$c_1^2(z) \not\equiv 0$ or $c_0^2(z) \not\equiv 0$
are included in
$\mathcal{K}_{1}^0, \mathcal{K}_{2}^0,
\mathcal{K}_{1}^2$ or $\mathcal{K}_{0}^2$,
which have already been determined in Prop \ref{det02}.
Thus, we only need to obtain
$\bar{\mathcal{K}}_2^1:=\mathcal{K}_2^1 
\cap \{ c_1^0(z)=c_2^0(z)=c_1^2(z)=c_0^2(z)\equiv 0 \}$,
$\bar{\mathcal{K}}_0^1:=\mathcal{K}_0^1 
\cap \{ c_1^0(z)=c_2^0(z)=c_1^2(z)=c_0^2(z)\equiv 0 \}$
to determine $\mathcal{K}_2^1, \mathcal{K}_0^1$.
It is easy to show
\begin{proposition}
\begin{align}
&(\mathrm{i}) \ \bar{\mathcal{K}}_{2}^1=\mathcal{B}_{\mathrm{I}}|_{
a_0=\bar{a}_0=a_1=\bar{a}_1=0, a_2 \neq 0} \cup
\mathcal{B}_{\mathrm{II}}|_{b_0=b_1=0, b_2 \neq 0}. \nn \\
&(\mathrm{ii}) \ \bar{\mathcal{K}}_{0}^1=\mathcal{B}_{\mathrm{I}}|_{
a_0=\bar{a}_0=a_1=\bar{a}_1=0, \bar{a}_2 \neq 0} \cup
\mathcal{B}_{\mathrm{II}}|_{b_0=b_1=0, b_5 \neq 0}. \nn
\end{align}
\label{det2101}
\hspace*{\fill}    \sq
\end{proposition}
It is enough to compute $\bar{\mathcal{K}}_2^1$ since
$\bar{\mathcal{K}}_0^1$ can be obtained from 
$\bar{\mathcal{K}}_2^1$ using \eqref{solutionspaceT}.
At last, we consider $\mathcal{D}=
\mathcal{D}^0 \cup \mathcal{D}^1 \cup \mathcal{D}^2$,
solutions  whose diagonal elements
are not identically 0. \\
Among $K(z) \in \mathcal{D}$,
any solution with some off-diagonal element
which is not identically 0 is included in
$\mathcal{K}_1^0, \mathcal{K}_2^0, \mathcal{K}_2^1, 
\mathcal{K}_1^2, \mathcal{K}_0^2$ or $\mathcal{K}_0^1$,
which have already been determined.
To obtain $\mathcal{D}$, what remains to be determined is
the solution space $\bar{\mathcal{D}}$, whose off-diagonal elements are all 0.
\begin{eqnarray}
\bar{\mathcal{D}}=\bar{\mathcal{D}^0} \cup 
\bar{\mathcal{D}^1} \cup \bar{\mathcal{D}^2}, \ \ \bar{\mathcal{D}^k}:=\mathcal{D}^k \cup \{ c_j^i(z) \not\equiv 0 \ \textrm{for} \ i \neq j \},
\ \ k=0,1,2. \nn
\end{eqnarray}
By a direct calculation, it is easy to show
\begin{proposition}
\begin{align}
\bar{\mathcal{D}}=
&\left\{
K(z)=\left(
\begin{array}{ccc}
c_{1}+c_{2}z^2 & 0 & 0 \\
0 & c_{2}z^2+c_{1}z^4 & 0 \\
0 & 0 & c_{2}z^2 +c_{1}z^4
\end{array}
\right)c(z) \ \Big| \
(c_{1}, c_{2}) \in \mathbb{P}^1(\mathbb{C})
\right\}
 \nn \\
&\bigcup \left\{
K(z)=\left(
\begin{array}{ccc}
c_{1}+c_{2}z^2 & 0 & 0 \\
0 & c_{1}+c_{2}z^2 & 0 \\
0 & 0 & c_{2}z^2 +c_{1}z^4
\end{array}
\right)c(z) \ \Big| \
(c_{1}, c_{2}) \in \mathbb{P}^1(\mathbb{C})
\right\}. \nn
\end{align}
\hspace*{\fill}  \sq
\label{detdiagonal}
\end{proposition}
From
Prop \ref{det02}, \ref{det2101} 
and \ref{detdiagonal}, $\mathcal{K}$ becomes
\\
\begin{align}
\mathcal{K}=&
\mathcal{K}_{1}^0 \bigcup \mathcal{K}_{1}^2 \bigcup 
\mathcal{K}_{2}^0 \bigcup \mathcal{K}_{0}^{2}
 \bigcup \mathcal{K}_{2}^{1} \bigcup 
\mathcal{K}_{0}^{1} \bigcup \mathcal{D} \nn \\
=&
\mathcal{K}_{1}^0 \bigcup \mathcal{K}_{1}^2 \bigcup 
\mathcal{K}_{2}^0 \bigcup \mathcal{K}_{0}^{2}
 \bigcup \bar{\mathcal{K}}_{2}^{1} \bigcup 
\bar{\mathcal{K}}_{0}^{1} \bigcup \bar{\mathcal{D}} \nn \\
=&
\mathcal{B}_{\mathrm{I}}|_{a_0 \neq 0} \bigcup 
\mathcal{B}_{\mathrm{I}}|_{\bar{a}_0 \neq 0}
\bigcup
\left\{\mathcal{B}_{\mathrm{I}}|_{a_1 \neq 0}  \bigcup
\mathcal{B}_{\mathrm{II}}|_{b_0 \neq 0}  \right\}
\bigcup
\left\{\mathcal{B}_{\mathrm{I}}|_{\bar{a}_1 \neq 0}  \bigcup
\mathcal{B}_{\mathrm{II}}|_{b_1 \neq 0}  \right\}
\nn \\
&\bigcup
\left\{\mathcal{B}_{\mathrm{I}}|_{a_0=\bar{a}_0=a_1=\bar{a}_1=0,
a_2 \neq 0}  \bigcup
\mathcal{B}_{\mathrm{II}}|_{b_0=b_1=0, b_2 \neq 0}  \right\}
\nn \\
&\bigcup
\left\{\mathcal{B}_{\mathrm{I}}|_{a_0=\bar{a}_0=a_1=\bar{a}_1=0,
\bar{a}_2 \neq 0}  \bigcup
\mathcal{B}_{\mathrm{II}}|_{b_0=b_1=0, b_5 \neq 0}  \right\}
\bigcup \bar{\mathcal{D}} \nn \\
=&
\left\{\mathcal{B}_{\mathrm{I}}|_{a_0 \neq  0 \ \textrm{or} \
\bar{a}_0 \neq  0 \ \textrm{or} \ a_1 \neq  0 \ \textrm{or} \
\bar{a}_1 \neq  0}  \bigcup
\mathcal{B}_{\mathrm{I}}|_{a_0=\bar{a}_0=a_1=\bar{a}_1=0, 
a_2 \neq 0 \ \textrm{or} \ \bar{a}_2 \neq 0}
\bigcup  \bar{\mathcal{D}}
\right\}
\nn \\
&\bigcup
\left\{\mathcal{B}_{\mathrm{II}}|_{b_0 \neq  0 \ \textrm{or} \
b_1 \neq  0}  \bigcup
\mathcal{B}_{\mathrm{II}}|_{b_0=b_1=0, 
b_2 \neq 0 \ \textrm{or} \ b_5 \neq 0}
\bigcup  \bar{\mathcal{D}}
\right\}. \label{comp1}
\end{align}
One can easily show the following Lemma about
$\bar{\mathcal{D}}$.
\begin{lemma}
\begin{align}
&(\mathrm{i}) \ \mathcal{B}_{\mathrm{I}}|_{a_0=\bar{a}_0=a_1=\bar{a}_1
=a_2=\bar{a}_2=0}=\bar{\mathcal{D}}, \nn \\
&(\mathrm{ii}) \ \mathcal{B}_{\mathrm{II}}|_{b_0=b_1=b_2=b_5=0}=\bar{\mathcal{D}}.
\nn
\end{align}
\label{identification}
\hspace*{\fill} \sq 
\end{lemma}
Thus, from \eqref{comp1}, Lemma \ref{identification}
and
\begin{align}
\mathcal{B}_{\mathrm{I}}=&
\mathcal{B}_{\mathrm{I}}|_{a_0 \neq  0 \ \textrm{or} \
\bar{a}_0 \neq  0 \ \textrm{or} \ a_1 \neq  0 \ \textrm{or} \
\bar{a}_1 \neq  0}  \bigcup
\mathcal{B}_{\mathrm{I}}|_{a_0=\bar{a}_0=a_1=\bar{a}_1=0, 
a_2 \neq 0 \ \textrm{or} \ \bar{a}_2 \neq 0} \nn \\
&\bigcup \mathcal{B}_{\mathrm{I}}|_{a_0=\bar{a}_0=a_1=\bar{a}_1=
a_2=\bar{a}_2=0}, \nn \\
\mathcal{B}_{\mathrm{II}}=&
\mathcal{B}_{\mathrm{II}}|_{b_0 \neq  0 \ \textrm{or} \
b_1 \neq  0}  \bigcup
\mathcal{B}_{\mathrm{II}}|_{b_0=b_1=0, 
b_2 \neq 0 \ \textrm{or} \ b_5 \neq 0}
\bigcup
\mathcal{B}_{\mathrm{II}}|_{b_0=b_1= 
b_2=b_5=0}, \nn
\end{align}
which obviously hold, one gets
\begin{align}
\mathcal{K}=\mathcal{B}_{\mathrm{I}} \cup
\mathcal{B}_{\mathrm{II}},
\end{align}
which proves Th \ref{subtheorem}.
\hspace*{\fill} \wsq \\
Investigating the algebraic varieties 
$\mathcal{V}_{\mathrm{I}}$ and $\mathcal{V}_{\mathrm{II}}$
which parametrize
$\mathcal{B}_{\mathrm{I}}$ and $\mathcal{B}_{\mathrm{II}}$,
one can show that
$\mathcal{B}_{\mathrm{I}}$ and $\mathcal{B}_{\mathrm{II}}$
can be described by spaces 
$\mathcal{A}_{\mathrm{I}}$ and $\mathcal{A}_{\mathrm{II}}$ (Def \ref{defspace}),
which are parametrized by simpler algebraic varieties
$\mathcal{U}_{\mathrm{I}}$ and $\mathcal{U}_{\mathrm{II}}$ (Def \ref{algebraicvariety}).
\begin{proposition}
\begin{align}
&(\mathrm{i}) \ \mathcal{B}_{\mathrm{I}}=\mathcal{A}_{\mathrm{I}}. \nn \\
&(\mathrm{ii}) \ \mathcal{B}_{\mathrm{II}}
=\mathcal{A}_{\mathrm{II}} \cup \{ K(z)=\mathrm{Id} \}. \nn
\end{align}
\hspace*{\fill} \sq
\label{spaceidentification}
\end{proposition}
The proof of Prop \ref{spaceidentification} (i) is given in Appendix C.
Combining Th \ref{subtheorem}, Prop \ref{spaceidentification} and 
\begin{align}
\{K(z)= \textrm{Id} \}=A_{\mathrm{I}}|_{D_2=E_1=E_2=0}, \nn
\end{align}
one has
\begin{equation}
\mathcal{K}=\mathcal{A}_{\mathrm{I}} \cup \mathcal{A}_{\mathrm{II}}. \nn
\end{equation}
Thus, the proof of Th \ref{maintheorem} is completed.
\hspace*{\fill} \wsq 
\section{Discussion}
In this paper, we considered the reflection equation of the
$N$=3 Cremmer-Gervais $R$-matrix.
The reflection equation is shown to be equivalent to
38 equations which do not depend on the parameters of the
$R$-matrix, $q$.
The solution space is determined by solving those 38 equations.
We found there are two types, each of which is parametrized by
the algebraic variety
$\mathbb{P}^1(\mathbb{C}) \times
\mathbb{P}^1(\mathbb{C}) \times
\mathbb{P}^2(\mathbb{C})$
and
$
\mathbb{C} \times
\mathbb{P}^1(\mathbb{C}) \times
\mathbb{P}^2(\mathbb{C})$.
The Cremmer-Gervais $R$-matrix satisfies the 
conservation law  \eqref{conservation}
and $T$-invariance \eqref{t-invariance}.
On the other hand, the critical $\mathbb{Z}_{n}$
vertex model satisfies the conservation law
and the $\mathbb{Z}_n$-invariance.
Since there does not exist a simple gauge tranformation
between these two $R$-matrices, we don't know the relation
between the $K$-matrices.
However, unexpectedly, the latter variety also appears in the solution of the
critical $\mathbb{Z}_{3}$
vertex model \cite{Y2}.
We mention that
one can also formulate and solve the dual reflection equation
\cite{Sklyanin}
for $R$-matrices not satisfying crossing unitarity \cite{DFIL}.
It is interesting to extend the analysis to
determine the full solution space
for the general $N$-state Cremmer-Gervais $R$-matrix.
Also interesting is to study the integrable model
associated with this $R$-matrix
under the periodic or open boundary condition.
\section*{Acknowledgments}
The authors thank A. Kuniba for helpful comment.
This work was supported in part by Global COE Program
(Global Center of Excellence for Physical Sciences Frontier),
MEXT, Japan.

\rnc{\theequation}{A.\arabic{equation}}\setcounter{equation}{0}
\rnc{\thesubsection}{A.\arabic{subsection}}\setcounter{subsection}{0}
\rnc{\thelemma}{A.\arabic{lemma}}\setcounter{lemma}{0}
\rnc{\theproposition}{A.\arabic{proposition}}\setcounter{proposition}{0}
\rnc{\thedefinition}{A.\arabic{definition}}\setcounter{definition}{0}

\section*{Appendix A. Proof of Prop \ref{step0}, Prop \ref{123123prime}
and \ref{123ABC}}
\subsection{Proof of Prop \ref{step0}}
We show 
Prop \ref{step0} by showing 2 Lemmas.
\begin{lemma}
\begin{equation}(00|11)=0, \ (00|22)=0. 
\end{equation}   \hspace*{\fill}   \sq
\label{step0no1}
\end{lemma}
{[} \textit{Proof of Lemma \ref{step0no1}} {]} \\
Directly calculating  $(00|ii)$ which is
an element of the reflection equation, one has
\begin{align}
(00|ii)=& \displaystyle \sum_{k_{1}, k_{2}, k_{3}, k_{4}=0}^{2}
{[} R_{12}(z_{1}/z_{2}) {]}_{k_{1} k_{2}}^{0 \ 0} {[}K_{1}(z_{1}) {]}_{k_{3}}^{k_{1}}
{[} R_{21}(z_{1} z_{2}) {]}_{i \ k_{4}}^{k_{3} k_{2}} {[} K_{2}(z_{2}) {]}_{i}^{k_{4}}
 \nn \\
&- \displaystyle \sum_{k_{1}, k_{2}, k_{3}, k_{4}=0}^{2}
{[} K_{2}(z_{2}) {]}_{k_{1}}^{0} 
{[}R_{12}(z_{1} z_{2})  {]}_{k_{2} k_{3}}^{0 \ k_{1}}
{[}K_{1}(z_{1}) {]}_{k_{4}}^{k_{2}}
{[} R_{21}(z_{1}/z_{2})  {]}_{i \ i}^{k_{4} k_{3}}
 \nn \\
=&{[}R_{12}(z_{1}/z_{2}){]}_{00}^{00} {[} K_{1}(z_{1}) {]}_{i}^0
{[} R_{21}(z_{1} z_{2})  {]}_{i0}^{i0} {[}K_{2}(z_{2})  {]}_{i}^0  \nn \\
&-{[} K_{2}(z_{2})  {]}_{i}^0 {[} R_{12}(z_{1} z_{2}) {]}_{0i}^{0i}
{[} K_{1}(z_{1}) {]}_{i}^0 {[} R_{21}(z_{1}/z_{2})  {]}_{ii}^{ii} \nn \\
=&\{ (q z_{2}/z_{1}-q^{-1}z_{1}/z_{2})/(q-q^{-1})(z_{1}/z_{2}-z_{2}/z_{1}) \}
c_{i}^0(z_{1}) \{ - q^{-1}/(q-q^{-1}) \} c_{i}^0(z_{2})
 \nn \\
&-c_{i}^0(z_{2})\{ - q^{-1}/(q-q^{-1}) \} c_{i}^0(z_{1})
\{ (q z_{2}/z_{1}-q^{-1}z_{1}/z_{2})/(q-q^{-1})(z_{1}/z_{2}-z_{2}/z_{1}) \}
 \nn \\
=&0. \nn
\end{align}
In the second equality, we picked up the terms which involve
$\[R(z) \]_{kl}^{ij} \not\equiv 0$.
\hspace*{\fill}  \wsq
\\
\begin{lemma}
\begin{equation}(22|11)=0, \ (22|00)=0. 
\end{equation}  \hspace*{\fill}      \sq 
\label{step0no2}
\end{lemma}
{[} \textit{Proof of Lemma \ref{step0no2}} {]} \\ 
Applying Lemma $ \ref{close}$ to
$(00|11)=0$ and $(00|22)=0$ shown in Lemma \ref{step0no1},
one has $T(00|11)=(22|11)=0$ and $T(00|22)=(22|00)=0$ respectively.
\hspace*{\fill} \wsq \\
Combining Lemma \ref{step0no1} and \ref{step0no2},
we have Prop \ref{step0}.  
\hspace*{\fill} \wsq
\subsection{Proof of Prop \ref{123123prime}}
We prove Prop \ref{123123prime} by several steps.
First, we show
\begin{lemma}
\begin{eqnarray}
{\bf 1}
\Longrightarrow {\bf 1^{\prime}}. \nn
\end{eqnarray}
   \hspace*{\fill}   \sq
\label{11prime}
\end{lemma}
{[} \textit{Proof of Lemma \ref{11prime}} {]} \\ 
This follows from
\begin{eqnarray} 
(00|21) &\Leftrightarrow& (00|12),  \\
(01|21), \ (02|22), \ (20|22) &\Rightarrow& 
(10|12), \ (00|20), \ (00|02).
\end{eqnarray}
Let us show
$(01|21), \ (02|22), \ (20|22) \Rightarrow (10|12)$ for example.
Calculating $(10|12)$, we find that it can be expressed using
$(01|21), (02|22)$ and $(20|22)$ as
\begin{align}
(10|12)=-q^2 (01|21)+(q^2-1)(q^2 z_2^2-z_1^2)^{-1}
(z_1^2(02|22)+ z_2^2(20|22)).
\end{align}
Since every element of
${\bf 1^{\prime}}$ can be expressed as combinations of elements of
${\bf 1}$, we have Lemma \ref{11prime}. \hspace*{\fill}  \wsq \\
Next we show
\begin{lemma}
\begin{eqnarray}
{\bf 1 \cup 2 \cup} T\!{\bf 2}
\Longrightarrow {\bf 1^{\prime} \cup 2^{\prime}}. \nn
\end{eqnarray}
\hspace*{\fill}    \sq
\label{1212prime}
\end{lemma}
{[} \textit{Proof of Lemma \ref{1212prime}} {]} \\ 
We first prove the following relations.
\begin{eqnarray} 
(00|21), \ (01|22) &\Rightarrow& (10|22),  \\
(02|11), \ (00|00), \ (12|12) &\Rightarrow& 
(21|21), \ (01|10), \\
(02|11), \ (00|00), \ (12|12), \ (10|10) &\Rightarrow& (11|11), \\
(02|11), \ (00|00), \ (10|10) &\Rightarrow& (20|20), \\
(01|12), \ (02|22), \ (20|22) &\Rightarrow& (11|22), (10|21).
\end{eqnarray}
For example,
$(02|11), \ (00|00),  \ (12|12) \Rightarrow (21|21)$
holds since $(21|21)$ can be expressed using
$(02|11)$, $(00|00)$ and $(12|12)$ as
\begin{align}
(21|21)=q^2(z_1^2-z_2^2)z_2^{-2} (02|11)
+z_1^4 z_2^2(q^2-1)(z_1^2-q^2 z_2^2)^{-1} (00|00)
-z_1 z_2^{-2}(12|12).
\end{align}
From these relations, one has
\begin{align}
{\bf 1} \cup {\bf 2} \cup (12|12) \Rightarrow {\bf 2^{\prime}},
\end{align}
which, combined with Lemma \ref{11prime}, shows Lemma \ref{1212prime}.
\hspace*{\fill} \wsq \\
Finally, let us show Prop \ref{123123prime}.
Let us notice the following relations holds.
\begin{eqnarray} 
(00|10), \ (20|21), \ (02|12) &\Rightarrow& 
(01|11), \ (00|01), \ (10|11)  \\
(00|10), \ (20|21), \ (20|12) &\Rightarrow& 
(02|21), \label{exampleeq} \\
\begin{array}{l} (00|10), \ (20|21), \ (02|12), \ (20|12) \\
(01|20), \ (21|22), \ (12|22)
\end{array} 
&\Rightarrow& (10|02), \\
\begin{array}{l} (01|02), \ (21|22), \ (12|22), \ (02|12) \\
(20|12), \ (00|10), \ (20|21)
\end{array} 
&\Rightarrow& (10|20), \ (11|12), \ (11|21). 
\end{eqnarray} 
For example, one can show \eqref{exampleeq} by noting the following equality
\begin{align}
(02|21)=-q^{-2}(20|12)+z_1^{-2}(z_2^2-z_1^2)(q^2-1)^{-1}(20|21)
+ z_2^{-2}(z_2^2-z_1^2)(q^2-1)^{-1}(02|12).
\end{align}
\hspace*{\fill} \wsq \\
The above relations mean
\begin{align}
{\bf 1} \cup {\bf 2} \cup {\bf 3} \Rightarrow {\bf 3^{\prime}}.
\end{align}
Combining this with Lemma \ref{1212prime},
one has Prop \ref{123123prime}.
\hspace*{\fill} \wsq 
\subsection{Proof of Prop \ref{123ABC}}
As is the case with Prop \ref{123123prime},
we show Prop \ref{123ABC} by several steps.
First, we have
\begin{lemma}
\begin{eqnarray}
{\bf 1}
\Longleftrightarrow {\bf A}. \nn
\end{eqnarray}
  \hspace*{\fill}   \sq
\label{1A}
\end{lemma}
{[} \textit{Proof of Lemma \ref{1A}} {]} \\
This follows from the relations below.
\begin{eqnarray} 
(02|11) &\Leftrightarrow& A_1,  \\
(00|21) &\Leftrightarrow& A_2,  \\
(02|12) &\Leftrightarrow& A_3,  \\
(01|21) &\Leftrightarrow& A_4,  \\
(00|00) &\Leftrightarrow&  A_5, \label{exampleequiv} \\
(02|20) &\Leftrightarrow&  A_6. \\
(02|22), \ (20|22) & \Leftrightarrow & A_7, \ A_8.
\label{exampleequivalence}
\end{eqnarray}
For example, calculating $(00|00)$, one has
\begin{align}
(00|00)=(q^2 z_2^2-z_1^2)(q^2-1)^{-1}(z_2^2-z_1^2)^{-1}(1-z_1^2 z_2^2)^{-1}A_5,
\end{align}
which shows \eqref{exampleequiv}. \\
Let us next show \eqref{exampleequivalence}.
We find by calculation that $(02|22)$ and $(20|22)$ can be expressed by 
$A_7$ and $A_8$ as
\begin{align}
(02|22)=(q^2-1)^{-1}(z_1^2-z_2^2)^{-1}(1-z_1^2 z_2^2)^{-1}(z_1^2 z_2^4 A_7+A_8),
\label{part1} \\
(20|22)=(q^2-1)^{-1}(z_1^2-z_2^2)^{-1}(z_1^2 z_2^2-1)^{-1}(z_1^4 z_2^2 A_7+q^2 A_8),
\label{part2}
\end{align}
from which we find 
\begin{eqnarray} 
(02|22), \ (20|22) & \Leftarrow & A_7, \ A_8.  \label{part3}
\end{eqnarray} 
Solving \eqref{part1} and \eqref{part2} for $A_7$ and $A_8$, one gets
\begin{eqnarray} 
(02|22), \ (20|22) & \Rightarrow & A_7, \ A_8,
\end{eqnarray} 
together with \eqref{part3}, shows \eqref{exampleequivalence}.
\hspace*{\fill} \wsq \\
The above relations imply the equivalence between 
the elements of the reflection equation in ${\bf 1}$ and
the relations among the matrix elements of $K(z)$ in
${\bf A}$, which is exactly
Lemma \ref{1A}. \hspace*{\fill}    \wsq \\
Next we prove
\begin{lemma}
\begin{eqnarray}
{\bf 1 \cup 2}
\Longleftrightarrow {\bf A \cup B}. \nn
\end{eqnarray}
  \hspace*{\fill}   \sq
\label{12AB}
\end{lemma}
{[} \textit{Proof of Lemma \ref{12AB}} {]} \\
The following relations can be shown.
\begin{eqnarray}
(20|21), \ (02|12) &\Rightarrow& B_1,  \\ 
(01|22), \ (00|21) &\Rightarrow & B_2,  \\
(01|12), \ (02|22), \ (20|22) &\Rightarrow& B_3, \\
(10|10), \ (00|00), \ (02|11) &\Rightarrow & B_4, \\
(12|21), \ (00|00), \ (02|11) &\Rightarrow & B_5, \\
(21|22), \ (12|22), \ (02|12) & \Rightarrow& B_6, \ B_7, \\
A_3, \ B_1 &\Rightarrow & (20|21),  \\
A_2, \ B_2 &\Rightarrow & (01|22),  \\
A_7, \ A_8, \ B_3 &\Rightarrow& (01|12),  \label{exampleshow} \\
A_1, \ A_5, \ B_4 & \Rightarrow & (10|10),  \\
A_1, \ A_5, \ B_5 & \Rightarrow & (12|21), \\
A_3, \ B_6, \ B_7 &\Rightarrow &(21|22), \ (12|22). 
\end{eqnarray} 
For example, \eqref{exampleshow} can be shown by
noting that $(01|12)$ can be expressed by 
$A_7, A_8$ and $B_3$ as
\begin{align}
(01|12)=(q^2-1)^{-1}(z_2^2-z_1^2)^{-1}(z_1^2 z_2^2)^{-1}
(A_8+z_1^2 z_2^4 A_7+z_2^2 (1-z_1^2 z_2^2) B_3).
\end{align}
From these relations, one has
\begin{align}
{\bf 1} \cup {\bf 2} \Longrightarrow {\bf B}, \\
{\bf A} \cup {\bf B} \Longrightarrow {\bf 2}. 
\end{align}
Combining these with Lemma \ref{1A}, we get
Lemma \ref{12AB}.
\hspace*{\fill} \wsq \\
Now let us finally prove Prop \ref{123ABC}.
One can show the following relations among the polynomials.
\begin{eqnarray}
(00|10), \ (20|21), \ (02|12) &\Rightarrow& C_1,  \\
(20|12), \ (00|10), \ (20|21), \ (02|12) &\Rightarrow& C_2,
 \\
(01|20), \ (02|12), \ (20|21), \ (21|22), \ (12|22)
&\Rightarrow & C_3,   \\
\begin{array}{l}
(01|02), \ (21|22), \ (12|22), \ (02|12) \\
(20|12), \ (00|10), \ (20|21)
\end{array}
& \Rightarrow &  C_4, \\
(11|02), \ (12|21), \ (02|11), \ (00|00)
& \Rightarrow &  C_5, \\
A_3, \ B_1, \ C_1 &\Rightarrow& (00|10), \\
A_3, \ B_1, \ C_1, \ C_2 &\Rightarrow& (20|12), \\
A_3, \ B_1, \ B_6, \ B_7, \ C_3  &\Rightarrow& (01|20), \label{complex} \\
B_6, \ C_2, \ C_4 & \Rightarrow& (01|02), \\
A_5, \ B_5, \ C_5 & \Rightarrow& (11|02).
\end{eqnarray}
For example, one can show \eqref{complex} since $(01|20)$ can be expressed as 
combinations of $A_3, B_1, B_6, B_7$ and $C_3$ as
\begin{align}
(01|20)=&(q^2-1)^{-1}(z_2^2-z_1^2)^{-1}(1-z_1^2 z_2^2)^{-1} \nn \\
&\times ((z_1^2+z_2^2)A_3-z_1^2(1-z_1^2 z_2^2)B_1+z_1^2 B_6-B_7+(1-z_1^2 z_2^2)C_3).
\end{align}
The above relations mean
\begin{align}
{\bf 1} \cup {\bf 2} \cup {\bf 3} \Longrightarrow {\bf C}, \\
{\bf A} \cup {\bf B} \cup {\bf C} \Longrightarrow {\bf 3},
\end{align}
which, combined with Lemma \ref{12AB}, shows Prop \ref{123ABC}.
\hspace*{\fill}    \wsq 

\rnc{\theequation}{B.\arabic{equation}}\setcounter{equation}{0}
\rnc{\thesubsection}{B.\arabic{subsection}}\setcounter{subsection}{0}
\rnc{\thelemma}{B.\arabic{lemma}}\setcounter{lemma}{0}
\rnc{\theproposition}{B.\arabic{proposition}}\setcounter{proposition}{0}
\rnc{\thedefinition}{B.\arabic{definition}}\setcounter{definition}{0}

\section*{Appendix B. Proof of Prop \ref{det02} (ii)}
We use the following simple Lemma
to calculate the solution to the reflection equation.
\begin{lemma}
If two meromorphic functions $X(z),Y(z) \in \mathcal{M}$
satisfy
\begin{equation}
X(z_{1})Y(z_{2})=X(z_{2})Y(z_{1}), \nn
\end{equation}
there exist constants $C_{1}, C_{2}$
and a meromorphic function $f(z) \in \mathcal{M}$
satisfying
\begin{equation}
X(z)=C_{1} f(z), \ Y(z)=C_{2} f(z). \nn 
\end{equation}
\label{lemma3} 
\hspace*{\fill}   \sq
\end{lemma}
We first use 8 equations 
$A_2=0, A_3=0, A_5^{\prime}=0, A_7=0,
A_8=0, B_2=0, B_3=0$ and $C_4=0$
out of 38 equations ${\bf Reduced^{\prime}}$
to prove the following.
\begin{proposition}
If $K(z) \in \mathcal{K}_{2}^0$, $K(z)$
must be expressed in the following form.
\begin{eqnarray}
&K(z)=((\alpha_4-\alpha_3)z^2-(\bar{\alpha}_4-\bar{\alpha}_3)z^4
+\alpha_7 z^6) \mathrm{Id} \, c(z) \nn \\
&+\left(
\begin{array}{ccc}
-\bar{\alpha}_{3} -\alpha_{7} z^2 
& \alpha_{0} & \alpha_{1}z^2 \\
\bar{\alpha}_{2}+\alpha_{5}z^2 & 0
&\alpha_{0}+\alpha_{2}z^2 \\
\bar{\alpha}_{1}+\alpha_{6}z^2 & \bar{\alpha}_{0}z^2 &
-\alpha_{3}z^2 
\end{array}
\right)(z^4-1)c(z), \label{(2)Kmatrix}
\end{eqnarray}
where $\alpha_{i}, i=0, \cdots ,7, \bar{\alpha}_{i}, i=0, \cdots ,4$ 
are constants, $\alpha_{1} \neq 0$, $c(z) \not\equiv 0$ is a meromorphic function,
and $\alpha_{i}, \bar{\alpha}_{i}$ satisfy 4 relations,
\begin{align}
\alpha_{0} \bar{\alpha}_{0}-\alpha_{1} \bar{\alpha}_{1}=0,
\ \alpha_{0} \alpha_{3}-\alpha_{1} \bar{\alpha}_{2}=0, 
\ \alpha_{0} \alpha_{2}-\bar{\alpha}_{3} \alpha_{1}=0, \
\alpha_{0}^2-\alpha_{1} \alpha_{4}=0.
\label{4relationstosatisfy}
\end{align}
 \hspace*{\fill}  \sq
\label{8equationsdet}
\end{proposition}
{[}\textit{Proof of Prop \ref{8equationsdet}} {]} \\
Since we are considering $K(z) \in \mathcal{K}_2^0$,
$c_2^0(z)$ can be expressed as
\begin{eqnarray}
c_{2}^0(z)= \alpha_{1} z^2(z^4-1)c(z), \label{(2)02} 
\end{eqnarray}
where $c(z) \not \equiv 0$ is a meromorphic function of $z$,
and $\alpha_{1} \neq 0 $ is a constant. \\
From $A_2=0$ and Lemma \ref{lemma3},
one can express $c_{1}^0(z)$ as
\begin{eqnarray}
c_{1}^0(z)= \alpha_{0} (z^4-1)c(z), \label{(2)01}
\end{eqnarray}
where $\alpha_{0}$ is a constant. \\
Similary, Utilizing $A_3=0, B_2=0, B_3=0$ and Lemma \ref{lemma3},
we get
\begin{align}
&c_{1}^2(z)= \bar{\alpha}_{0}z^2(z^4-1)c(z), \label{(2)21} \\
&c_{2}^1(z)=(\alpha_{0}+ \alpha_{2} z^2)(z^4-1)c(z), \label{(2)12} \\
&c_{2}^2(z)=c_{1}^1(z)-\alpha_{3}z^2(z^4-1)c(z), \label{(2)2211}
\end{align}
where $\bar{\alpha}_{0}, \alpha_{2}, \alpha_{3}$ are constants. \\
Next, we substitute \eqref{(2)02} $\sim$ \eqref{(2)2211} into
$A_7=0, A_5^{\prime}=0, C_4=0$ and use Lemma \ref{lemma3} to obtain
\begin{align}
& c_{0}^0(z)=c_{1}^1(z)-\Big( \frac{\alpha_{0} \alpha_{2}}{\alpha_{1}}
+\alpha_{7}z^2 \Big)(z^4-1)c(z), \label{(2)0011} \\
& c_{0}^1(z)= \Big( \frac{\alpha_{0} \alpha_{3}}{\alpha_{1}}+\alpha_{5}z^2 \Big)
(z^4-1)c(z),  \label{(2)10}  \\
& c_{0}^2(z)=\Big( \frac{\alpha_{0} \bar{\alpha}_{0}}{\alpha_{1}}
+\alpha_{6}z^2 \Big)
(z^4-1)c(z), \label{(2)20} 
\end{align}
where $\alpha_{5}, \alpha_{6}, \alpha_{7}$ are constants.
We set
$\bar{\alpha}_{1}:=\alpha_{0} \bar{\alpha}_{0}/\alpha_{1},
\bar{\alpha}_{2}:=\alpha_{0} \alpha_{3}/\alpha_{1},
\bar{\alpha}_{3}:=\alpha_{0} \alpha_{2}/\alpha_{1}$. \\
Furthermore, we substitute \eqref{(2)02} $\sim$ \eqref{(2)20}
in $A_8=0$. The result is
\begin{eqnarray}
&&(z_{1}^4-1)(z_{2}^4-1)(z_{2}^4 c(z_{2})
( \alpha_{1} c_{1}^1(z_{1})+(( \alpha_{1} \alpha_{3}-\alpha_{0}^2)z_{1}^2
-\alpha_{1} \alpha_{7} z_{1}^6)c(z_{1})) \nn \\
&&-( \alpha_{1} c_{1}^1(z_{2})+(( \alpha_{1} \alpha_{3}-\alpha_{0}^2)z_{2}^2
-\alpha_{1} \alpha_{7} z_{2}^6)c(z_{2}))z_{1}^4 c(z_{1}))=0. \nn
\end{eqnarray}
Using Lemma \ref{lemma3}, one has
\begin{equation}
c_{1}^1(z)=\Big( \Big( \frac{\alpha_{0}^2}{\alpha_{1}}-\alpha_{3} \Big)z^2
+\bar{\alpha}_{4}^{\prime}z^4 +\alpha_{7}z^6 \Big)c(z),  \label{(2)11first}
\end{equation}
where $\bar{\alpha}_{4}^{\prime}$ is a constant.
Setting
$\alpha_{4}:= \alpha_{0}^2/\alpha_{1}$,
$\bar{\alpha}_{4}:=-\bar{\alpha}_{4}^{\prime}
+\bar{\alpha}_{3}$, \eqref{(2)11first} becomes
\begin{equation}
c_{1}^1(z)=((\alpha_{4}-\alpha_{3})z^2-(\bar{\alpha}_{4}-\bar{\alpha}_{3})z^4
+\alpha_{7} z^6)c(z). \label{(2)11}
\end{equation}
Substituting
\eqref{(2)11} into \eqref{(2)2211}, \eqref{(2)0011},
one sees $c_{2}^2(z)$ and $c_{0}^0(z)$
are expressed as
\begin{eqnarray}
&& c_{2}^2(z)=(\alpha_{4}z^2-(\bar{\alpha}_{4}-\bar{\alpha}_{3})z^4
+(\alpha_{7}-\alpha_{3})z^6 )c(z), \label{(2)22} \\
&& c_{0}^0(z)=(\bar{\alpha}_{3}+(\alpha_{4}+\alpha_{7}-\alpha_{3})z^2 
-\bar{\alpha}_{4} z^4)c(z). \label{(2)00} 
\end{eqnarray}
From
\eqref{(2)02}, \eqref{(2)01},
\eqref{(2)21}, \eqref{(2)12},
\eqref{(2)10}, \eqref{(2)20},
\eqref{(2)11}, \eqref{(2)22}, \eqref{(2)00}
and
$\bar{\alpha}_{1}:=\alpha_{0} \bar{\alpha}_{0}/\alpha_{1}$,
$\bar{\alpha}_{2}:=\alpha_{0} \alpha_{3}/\alpha_{1},
\bar{\alpha}_{3}:=\alpha_{0} \alpha_{2}/\alpha_{1}$,
$\alpha_{4}:=\alpha_{0}^2/\alpha_{1}$,
one sees that if
$K(z) \in \mathcal{K}_2^0$,
the matrix elements $c_j^i(z), i,j=0,1,2$
of $K(z)$ must be expressed as \eqref{(2)Kmatrix},
and $\alpha_i, \bar{\alpha}_i$ should satisfy
\eqref{4relationstosatisfy}.
Thus, Proposition \ref{8equationsdet} is proved.
\hspace*{\fill} \wsq  \\
Proposition \ref{8equationsdet} is just a necessary condition
to be a solution since we only used 8 equations of the
38 equations ${\bf Reduced^{\prime}}$.
For \eqref{(2)Kmatrix} to be a solution,
the rest of the 30 equations, into which 
\eqref{(2)Kmatrix} have been substituted,
must hold for any $z$.
These condtitions are
relations between the coefficients.
For example,
$T\!A_2=0$ becomes
\begin{equation}
\alpha_{6} \bar{\alpha}_{0}z_{1}^2 z_{2}^2(z_{1}^2-z_{2}^2)
(z_{1}^4-1)(z_{2}^4-1)c(z_{1})c(z_{2})=0. \label{need2}
\end{equation}
One must have
\begin{equation}
\alpha_{6} \bar{\alpha}_{0}=0, \label{(2)no1}
\end{equation}
for \eqref{need2} to hold for any $z$.
Similarly, the following relations must be fulfilled.
We denote the equations which yield the relations
to the right of them.
\begin{align}
\alpha_{6} \alpha_{0}=0, & \ T\!A_3=0, T\!C_1=0, \label{(2)no2} \\
(\bar{\alpha}_{0}-\alpha_{5}) \bar{\alpha}_{0}=0,    & \ T\!A_4=0, 
\label{(2)no3} \\
\bar{\alpha}_{2} \alpha_{0}-\alpha_{3} \alpha_{4}
-(\alpha_{2} \bar{\alpha}_{0}-\bar{\alpha}_{3} \bar{\alpha}_{4})
+\alpha_{7} \alpha_{4}=0, & \ A_6^{\prime}=0,  \label{(2)no22}  \\
\bar{\alpha}_{0} \bar{\alpha}_{2}-\alpha_{3} \bar{\alpha}_{1}
+\alpha_{7} \bar{\alpha}_{1}-\alpha_{6} \bar{\alpha}_{3}=0,
& \ T\!A_7=0, \label{(2)no5} \\
\bar{\alpha}_{0} \alpha_{5}-\bar{\alpha}_{1} \bar{\alpha}_{4}
-\alpha_{6} \alpha_{4}=0, & \ T\!A_8=0, \label{(2)no6} \\
\alpha_{7} \alpha_{0}=0, & \ B_1=0, T\!B_7=0, 
\label{(2)no7} \\
(\bar{\alpha}_{0}-\alpha_{5}) \bar{\alpha}_{1}+\alpha_{6} \bar{\alpha}_{2}=0,
& \ T\!B_2=0, \label{(2)no8} \\
\alpha_{7} \bar{\alpha}_{1}- \alpha_{6} \bar{\alpha}_{3}=0, & \
T\!B_3=0, \label{(2)no9} \\
(\bar{\alpha}_{0}-\alpha_{5}) \alpha_{0}=0, & \ B_4=0, 
\label{(2)no13} \\
\bar{\alpha}_{2} \alpha_{0}-\alpha_{3} \alpha_{4}=0, & \
B_5=0, \label{(2)no11} \\
\alpha_{7} \alpha_{3}-\alpha_{6} \alpha_{1}=0, & \
B_5=0, \label{(2)no12} \\
\alpha_{2} \bar{\alpha}_{0}-\bar{\alpha}_{3} \bar{\alpha}_{4}
+\alpha_{7}(\alpha_{3}-\alpha_{4})-\alpha_{6} \alpha_{1}=0, & \
T\!B_5=0, \label{(2)no10} \\
\alpha_{5} \bar{\alpha}_{3}-\bar{\alpha}_{1} \alpha_{2}
+\alpha_{6} \alpha_{0}-\alpha_{7} \bar{\alpha}_{2}=0,
& \ T\!B_6=0, \label{(2)no14} \\
\alpha_{2} \alpha_{4}-\alpha_{0} \bar{\alpha}_{3}-
( \alpha_{1} \bar{\alpha}_{0}-\alpha_{0} \bar{\alpha}_{4})=0,
& \ B_7=0, \label{(2)no16} \\
(\bar{\alpha}_{0}-\alpha_{5}) \alpha_{1}-\alpha_{7} \alpha_{2}=0,
& \ B_7=0, C_3=0, \label{(2)no17} \\
\bar{\alpha}_{2} \bar{\alpha}_{4}- \bar{\alpha}_{0} \alpha_{3}
-( \bar{\alpha}_{1} \alpha_{0}-\alpha_{5} \alpha_{4}
)+\alpha_{6} \alpha_{2}=0, & \ T\!B_7=0, \label{(2)no15} \\
\alpha_{1} \bar{\alpha}_{0}-\alpha_{0} \bar{\alpha}_{4}=0,
& \ C_1=0, \label{(2)no20} \\
\bar{\alpha}_{1} \alpha_{0}-\bar{\alpha}_{0} \alpha_{4}=0, & \ 
T\!C_1=0, \label{(2)no18} \\
\alpha_{7} \bar{\alpha}_{0}=0, & \
T\!C_1=0, T\!C_2=0, \label{(2)no19} \\
(\bar{\alpha}_{0}-\alpha_{5}) \alpha_{3} - \alpha_{6} \alpha_{2}=0, & \ 
T\!C_3=0, \label{(2)no21}  \\
\alpha_{0} \bar{\alpha}_{3}- \alpha_{2} \bar{\alpha}_{1}+\alpha_{6} \alpha_{0}=0,
& \ T\!C_4=0, \label{(2)no23} \\
\alpha_2 \bar{\alpha}_{2}-\alpha_{3} \bar{\alpha}_{3}=0,
& \ C_5^{\prime}=0, \label{(2)no25} \\ 
(\bar{\alpha}_{0}-\bar{\alpha}_{5}) 
\alpha_{0}+\alpha_{2} \bar{\alpha}_{2}-\alpha_{3} \bar{\alpha}_{3}=0,
& \ T\!C_5^{\prime}=0. \label{(2)no24} 
\end{align}
From Prop \ref{8equationsdet}, we also have
\begin{align}
\alpha_{0} \bar{\alpha}_{0}- \alpha_{1} \bar{\alpha}_{1}&=0, 
\label{(2)no26} \\
\alpha_{0} \alpha_{3}- \alpha_{1} \bar{\alpha}_{2}&=0,
\label{(2)no27} \\
\alpha_{0} \alpha_{2}-\bar{\alpha}_{3} \alpha_{1}&=0, 
\label{(2)no28} \\
\alpha_{0}^2- \alpha_{1} \alpha_{4}&=0. \label{(2)no29} 
\end{align}
Among
\eqref{(2)no1}$\sim$ \eqref{(2)no29}, we first 
consider the following 6 relations.
\begin{eqnarray}
\left\{
\begin{array}{ccc}
\alpha_{6} \bar{\alpha}_{0}=0 \ \eqref{(2)no1}, & \alpha_{6} \alpha_{0}=0 
\ \eqref{(2)no2}, & (\bar{\alpha}_{0}-\alpha_{5}) \bar{\alpha}_{0}=0 \
\eqref{(2)no3}, \nn \\
(\bar{\alpha}_{0}-\alpha_{5}) \alpha_{0}=0 \ \eqref{(2)no13},
& \alpha_{7} \bar{\alpha}_{0}=0 \ \eqref{(2)no19},
& \alpha_{7} \alpha_{0}=0 \ \eqref{(2)no7}
\end{array}
\right\}.
\end{eqnarray}
These 6 relations are equivalent to
\begin{equation}
(A) \{ \alpha_{6}=\alpha_{7}=0, \alpha_{5}= \bar{\alpha}_{0} \} \ \
\textrm{or} \ \ 
(B) \{ \alpha_{0}= \bar{\alpha}_{0}=0 \}. \nn
\end{equation}
Let us investigate the case
$(A)$ and $(B)$, separately.
\\
$(A) \ \alpha_{6}= \alpha_{7}=0, \alpha_{5}= \bar{\alpha}_{0}$  \\
Substituting
$\alpha_{6}= \alpha_{7}=0, \alpha_{5}= \bar{\alpha}_{0}$,
the matrix elements of $K(z)$ \eqref{(2)Kmatrix} become
\begin{align}
&K(z) \nn \\
&=\left(
\begin{array}{ccc}
\bar{\alpha}_{3}+(\alpha_{4}-\alpha_{3})z^2 -\bar{\alpha}_{4} z^4 & \alpha_{0}(z^4-1) 
& \alpha_{1}z^2(z^4-1) \\
(\bar{\alpha}_{2}+\bar{\alpha}_{0}z^2)(z^4-1) &
(\alpha_{4}-\alpha_{3})z^2-(\bar{\alpha}_{4}-\bar{\alpha}_{3})z^4 & 
(\alpha_{0}+\alpha_{2}z^2)(z^4-1) \\
\bar{\alpha}_{1}(z^4-1) & \bar{\alpha}_{0}z^2(z^4-1) &
\alpha_{4}z^2-(\bar{\alpha}_{4}-\bar{\alpha}_{3})z^4-\alpha_{3}z^6 
\end{array}
\right)c(z),  \label{(1)Kmatrix}
\end{align}
and the relations \eqref{(2)no1} $\sim$ \eqref{(2)no29} are reduced to
\begin{align}
&\alpha_{0} \bar{\alpha}_{0}-\alpha_{1} \bar{\alpha}_{1}=0, \label{(1)no1} \\
&\bar{\alpha}_{2} \alpha_{0}-\alpha_{3} \alpha_{4}
-(\alpha_{2} \bar{\alpha}_{0}-\bar{\alpha}_{3} \bar{\alpha}_{4})=0,
\label{(1)no13} \\
&\alpha_{0} \alpha_{2}-\bar{\alpha}_{3} \alpha_{1}=0,
\label{(1)no3} \\
&\bar{\alpha}_{0} \bar{\alpha}_{2}-\alpha_{3} \bar{\alpha}_{1}=0,
\label{(1)no2} \\
&\alpha_{0}^2-\alpha_{1} \alpha_{4}=0, \label{(1)no5} \\
&\bar{\alpha}_{0}^2-\bar{\alpha}_{1} \bar{\alpha}_{4}=0, 
\label{(1)no4} \\
&\bar{\alpha}_{2} \alpha_{0}-\alpha_{3} \alpha_{4}=0,
\label{(1)no7} \\
&\alpha_{2} \bar{\alpha}_{0}-\bar{\alpha}_{3} \bar{\alpha}_{4}=0, \label{(1)no6} \\
&\alpha_{0} \alpha_{3}-\alpha_{1} \bar{\alpha}_{2}=0,
\label{(1)no9} \\
&\bar{\alpha}_{0} \bar{\alpha}_{3}-\bar{\alpha}_{1} \alpha_{2}=0, \label{(1)no8} \\
&\alpha_{1} \bar{\alpha}_{0}-\alpha_{0} \bar{\alpha}_{4}
-(\alpha_{2} \alpha_{4}-\alpha_{0} \bar{\alpha}_{3})=0,
\label{(1)no11} \\
&\bar{\alpha}_{1} \alpha_{0}-\bar{\alpha}_{0} \alpha_{4}
-(\bar{\alpha}_{2} \bar{\alpha}_{4}-\bar{\alpha}_{0} \alpha_{3})=0, \label{(1)no10} \\
&\bar{\alpha}_{1} \alpha_{0}-\bar{\alpha}_{0} \alpha_{4}=0, \label{(1)no12} \\
&\alpha_{1} \bar{\alpha}_{0}-\alpha_{0} \bar{\alpha}_{4}=0, \label{(1)no15} \\
&\alpha_{2} \bar{\alpha}_{2}-\alpha_{3} \bar{\alpha}_{3}=0.
\label{(1)no14}
\end{align}
Noting that 
\eqref{(1)no13} can be obtained by subtracting the both hand sides of
\eqref{(1)no7} by those of \eqref{(1)no6}, the 15 relations
\eqref{(1)no1}$ \sim $ \eqref{(1)no15} are equivalent to 14 relations
\eqref{relation1} $(\mathrm{I}_j,j=1, \cdots , 8),
(T\mathrm{I}_j, j=3, \cdots , 8)$. 
Thus, in the case $(A)$,
the matrix elements of the solution
$K(z)$ is given by \eqref{(1)Kmatrix}, 
and the coefficients must satisfy the relations
$(\mathrm{I}_j,j=1, \cdots , 8), (T\mathrm{I}_j,j=3, \cdots , 8)$.
This corresponds to $\mathcal{B}_{\mathrm{I}}$ in
Def \ref{spacedef3} with $a_1 \neq 0$.
\\
$(B) \ \alpha_{0}=\bar{\alpha}_{0}=0$ \\
From 
\eqref{(2)no26}, \eqref{(2)no27}, \eqref{(2)no28}, \eqref{(2)no29}
and $\alpha_{1} \neq 0$, one must have
$\bar{\alpha}_{1}=\bar{\alpha}_{2}=\bar{\alpha}_{3}=\alpha_{4}=0$.
Thus the matrix elements of $K(z)$ \eqref{(2)Kmatrix} become
\begin{eqnarray}
K(z)=\left(
\begin{array}{ccc}
\alpha_{7}-\alpha_{3} -\bar{\alpha}_{4} z^2 
& 0 & \alpha_{1}(z^4-1) \\
\alpha_{5}(z^4-1) &
-\alpha_{3}-\bar{\alpha}_{4}z^2+\alpha_{7} z^4
&\alpha_{2}(z^4-1) \\
\alpha_{6}(z^4-1) & 0 &
-\bar{\alpha}_{4}z^2+(\alpha_{7}-\alpha_{3})z^4
\end{array}
\right)z^2c(z), \label{secondKmatrix}
\end{eqnarray}
and the relations
\eqref{(2)no1}$\sim$ \eqref{(2)no29} are reduced to 3 relations
\begin{align}
\alpha_{1} \alpha_{6}-\alpha_{7} \alpha_{3} &=0,  \label{equation2-1}   \\ 
\alpha_{6} \alpha_{2}+\alpha_{3} \alpha_{5} &=0,  \label{equation2-2}   \\ 
\alpha_{2} \alpha_{7}+\alpha_{5} \alpha_{1} &=0.  \label{equation2-3}
\end{align}
In the case $(B)$,
the matrix elements of the solution
$K(z)$ is given by \eqref{secondKmatrix}, 
and the coefficients must satisfy the relations
\eqref{equation2-1}, \eqref{equation2-2}, \eqref{equation2-3}.
This corresponds to $\mathcal{B}_{\mathrm{II}}$ in
Def \ref{spacedef3} with $b_0 \neq 0$. \\
Investigating the two cases $(A)$ and $(B)$,
one has
$\mathcal{K}_{2}^0=\mathcal{B}_{\mathrm{I}}|_{a_1 \neq 0} \cup \mathcal{B}_{\mathrm{II}}|_{b_0 \neq 0}$.
Thus, Prop \ref{det02} (ii) is proved.
\hspace*{\fill}  \wsq 

\rnc{\theequation}{C.\arabic{equation}}\setcounter{equation}{0}
\rnc{\thesubsection}{C.\arabic{subsection}}\setcounter{subsection}{0}
\rnc{\thelemma}{C.\arabic{lemma}}\setcounter{lemma}{0}
\rnc{\theproposition}{C.\arabic{proposition}}\setcounter{proposition}{0}
\rnc{\thedefinition}{C.\arabic{definition}}\setcounter{definition}{0}

\section*{Appendix C. Proof of Prop \ref{spaceidentification} (i)}
Observing the algebraic variety $\mathcal{V}_{\mathrm{I}}$ which parametrizes
$\mathcal{B}_{\mathrm{I}}$,
one notices the following Lemma holds.
\begin{lemma}
The points in $\mathcal{V}_{\mathrm{I}}$ 
fulfill the equation
\begin{align}
\mathrm{I}_0: \ a_0 \bar{a}_0-a_4 \bar{a}_4=0,
\end{align}
if any one of
$a_0, \bar{a}_0, a_1, \bar{a}_1, a_2, \bar{a}_2, a_3$ or $\bar{a}_3$
is nonzero.
\label{lemmacondition}
\hspace*{\fill} \sq
\end{lemma}
The 15 equations
$(\mathrm{I}_j, j=1, \cdots 8), (T \mathrm{I}_j, j=3, \cdots 8)$
and $\mathrm{I}_0$
can be conveniently expressed as
\begin{align}
\textrm{rank} \left|
\begin{array}{cccccc}
a_0 & a_1 & a_2 & a_3 & \bar{a}_0 & \bar{a}_4  \\
a_4 & a_0 & \bar{a}_3 & \bar{a}_2 & \bar{a}_1 & \bar{a}_0  
\end{array}
\right| =1. \nn
\end{align}
By utilizing Lemma \ref{lemmacondition}, the following holds.
\begin{proposition}
\begin{eqnarray}
&\mathcal{V}_{\mathrm{I}}=
\mathcal{V}_{\mathrm{I}}^0
\sqcup
\mathcal{V}_{\mathrm{I}}^1, \nn
\end{eqnarray}
where
\begin{align}
\mathcal{V}_{\mathrm{I}}^0&=\left\{
\left(
\begin{array}{ccccc}
0 & 0 & 0 & 0 & a_4 \\
0 & 0 & 0 & 0 & \bar{a}_4
\end{array}
\right) \in \mathbb{P}^9(\mathbb{C}) \ \Big| \
(a_4, \bar{a}_4) \in \mathbb{P}^1(\mathbb{C}), \
a_4 \bar{a}_4 \neq 0 \right\}, \nn \\
\mathcal{V}_{\mathrm{I}}^1&=\left\{
\left(
\begin{array}{ccccc}
a_0 & a_1 & a_2 & a_3 & a_4 \\
\bar{a}_0 & \bar{a}_1 & 
\bar{a}_2 & \bar{a}_{3} & \bar{a}_4 
\end{array}
\right) \in \mathbb{P}^9(\mathbb{C}) \ \Big| \
\mathrm{rank} \left|
\begin{array}{cccccc}
a_0 & a_1 & a_2 & a_3 & \bar{a}_0 & \bar{a}_4  \\
a_4 & a_0 & \bar{a}_3 & \bar{a}_2 & \bar{a}_1 & \bar{a}_0  
\end{array}
\right| =1 \right\}. \nn
\end{align}
\hspace*{\fill} \sq
\end{proposition}
Studying $\mathcal{V}_{\mathrm{I}}^1$ which is the main part of
$\mathcal{V}_{\mathrm{I}}$,
we find that the following variety is included.
\begin{definition}
\begin{eqnarray}
\mathcal{S}=\left\{
(c_{0}, c_{1}, c_{2}, c_{3}, c_{4}, c_{5})
\in \mathbb{P}^5 (\mathbb{C}) \ \Big| \
\begin{array}{l}
c_{j}, \ j=0 \sim 5
 \ \textrm{satisfy} \ 3 \ \textrm{relations in}  
\ \eqref{segrerelation} 
 \\
\mathcal{S}_1, \mathcal{S}_2, \mathcal{S}_3 
\end{array}
\right\}, 
\label{segre}
\end{eqnarray}
\begin{eqnarray}
\begin{array}{cccccc}
\mathcal{S}_1: \ c_{0} c_{1}-c_{3} c_{4}=0, &
\mathcal{S}_2: \ c_{1} c_{2}-c_{4} c_{5}=0, &
\mathcal{S}_3: \ c_{2} c_{3}-c_{5} c_{0}=0.
\end{array}
\label{segrerelation}
\end{eqnarray} \hspace*{\fill} \sq
\label{threefold}
\end{definition}
The coordinates of points of $\mathcal{V}_{\mathrm{I}}^1$
can be parametrized using the projective varieties $\mathbb{P}^1(\mathbb{C})$
and $\mathcal{S}$.
\begin{proposition}
\begin{eqnarray}
\mathcal{V}_{\mathrm{I}}^1=\mathcal{W}, \nn
\end{eqnarray}
where
\begin{align}
\mathcal{W}&=\left\{
\left(
\begin{array}{ccccc}
A_1 A_2 & A_1^2 & B_2 \bar{B}_2 & B_1 \bar{B}_2 & A_2^2 \\
\bar{A}_{1} \bar{A}_2 & \bar{A}_1^2 & 
B_{1} \bar{B}_1 & \bar{B}_{1} B_2 & \bar{A}_{2}^2 
\end{array}
\right) \in \mathbb{P}^9(\mathbb{C}) \ \Big| \
\begin{array}{l}
(B_1, B_2) \cong \mathbb{P}^1(\mathbb{C}), \\
(A_1, \bar{A}_1, \bar{B}_2, A_2, \bar{A}_2, \bar{B}_1) \in
\mathcal{S}
\end{array} 
\right\}. \nn
\end{align}
\hspace*{\fill} \sq
\label{point}
\end{proposition}
{[} \textit{Proof of Prop \ref{point}} {]} \\
Let us first take a look at 3 equations $\mathrm{I}_2, \mathrm{I}_3, T\!\mathrm{I}_3$
out of the 15 ones
$(\mathrm{I}_j, \ j=0, \cdots , 8), \
(T\mathrm{I}_j, \  j=3, \cdots , 8 ) $.
From $\mathrm{I}_2$, $\mathrm{I}_3$ and $T\!\mathrm{I}_3$,
we see that $a_j, \bar{a}_j, j=0, \cdots ,4$
can be parametrized as
\begin{align}
a_2=B_2 \bar{B}_2, \ \bar{a}_2=B_1 \bar{B}_1, \ a_3=B_1 \bar{B}_2, \ 
\bar{a}_3=\bar{B}_1 B_2, \nn
\end{align}
\begin{align}
a_0=A_1 A_2,  \ a_1=A_1^2, \ a_4=A_2^2, \nn
\end{align}
and
\begin{align}
\bar{a}_0=\bar{A}_1 \bar{A}_2, \ \bar{a}_1=\bar{A}_1^2, \ \bar{a}_4=\bar{A}_2^2, \nn
\end{align}
where $A_1, A_2, B_1, B_2, \bar{A}_1, \bar{A}_2, \bar{B}_1, \bar{B}_2 \in
\mathbb{C}$. \\
Substituting these parametrization into the remaining 12 equations among
$(\mathrm{I}_j, \ j=0, \cdots , 8), \
(T\mathrm{I}_j, \  j=3, \cdots , 8 ) $ and calculating 
$\mathcal{V}_{\mathrm{I}}^1|_{a_1 \neq 0},
\mathcal{V}_{\mathrm{I}}^1|_{\bar{a}_1 \neq 0},
\mathcal{V}_{\mathrm{I}}^1|_{a_1=\bar{a}_1=0, a_4 \neq 0},
\mathcal{V}_{\mathrm{I}}^1|_{a_1=\bar{a}_1=0, \bar{a}_4 \neq 0}
$ and
$\mathcal{V}_{\mathrm{I}}^1|_{a_1=\bar{a}_1=a_4=\bar{a}_4=0}$,
one finds
\begin{align}
\mathcal{V}_{\mathrm{I}}^1|_{a_1 \neq 0}=&\left\{
\left(
\begin{array}{ccccc}
A_1 A_2 & A_1^2 & B_2 \bar{B}_2 & B_1 \bar{B}_2 & A_2^2 \\
\bar{A}_{1} \bar{A}_2 & \bar{A}_1^2 & 
B_{1} \bar{B}_1 & \bar{B}_{1} B_2 & \bar{A}_{2}^2 
\end{array}
\right) \in \mathbb{P}^9(\mathbb{C}) \ \Big| \
\begin{array}{l}
A_1 \in \mathbb{C}^{\times}, \ (B_1, B_2) \in \mathbb{P}^1(\mathbb{C}),
\\
(A_1, \bar{A}_1, \bar{B}_2, A_2, \bar{A}_2, \bar{B}_1) \in
\mathcal{S}
\end{array} 
\right\}  \nn \\
=&\mathcal{W}|_{A_1 \neq 0}, \label{solutioncase1} \\
\mathcal{V}_{\mathrm{I}}^1|_{\bar{a}_1 \neq 0}
=&\mathcal{W}|_{\bar{A}_1 \neq 0}, \label{solutioncase2} \\
\mathcal{V}_{\mathrm{I}}^1|_{a_1=\bar{a}_1=0, a_4 \neq 0}=&
\left\{
\left(
\begin{array}{ccccc}
0 & 0 & 0 & 0 & A_2^2 \\
0 & 0 & 
B_{1} \bar{B}_1 & \bar{B}_{1} B_2 & 0
\end{array}
\right) \in \mathbb{P}^9(\mathbb{C}) \ \Big| \
A_2 \in \mathbb{C}^{\times}, \ \bar{B}_1 \in \mathbb{C}, \
(B_1, B_2) \in \mathbb{P}^1(\mathbb{C})
\right\} \nn \\
=&\mathcal{W}|_{A_1=\bar{A}_1=0, A_2 \neq 0},
\label{solutioncase3} \\
\mathcal{V}_{\mathrm{I}}^1|_{a_1=\bar{a}_1=0, \bar{a}_4 \neq 0}
=&\mathcal{W}|_{A_1=\bar{A}_1=0, \bar{A}_2 \neq 0},
\label{solutioncase4} \\
\mathcal{V}_{\mathrm{I}}^1|_{a_1=\bar{a}_1=a_4=\bar{a}_4=0}=&\left\{
\left(
\begin{array}{ccccc}
0 & 0 & B_2 \bar{B}_2 & B_1 \bar{B}_2 & 0 \\
0 & 0 & B_{1} \bar{B}_1 & \bar{B}_{1} B_2 & 0 
\end{array}
\right) \in \mathbb{P}^9(\mathbb{C}) \ \Big| \
(B_1, B_2) \in \mathbb{P}^1(\mathbb{C}), \
(\bar{B}_1, \bar{B}_2) \in \mathbb{P}^1(\mathbb{C})
\right\} \nn \\
=&\mathcal{W}|_{A_1=\bar{A}_1=A_2=\bar{A}_2=0}. 
\label{solutioncase5}
\end{align}
From
\eqref{solutioncase1}, \eqref{solutioncase2}, \eqref{solutioncase3},
\eqref{solutioncase4}, \eqref{solutioncase5} and the following
obvious relations
\begin{align}
\mathcal{V}_{\mathrm{I}}^1&=\mathcal{V}_{\mathrm{I}}^1|_{a_1 \neq 0} \cup
\mathcal{V}_{\mathrm{I}}^1|_{\bar{a}_1 \neq 0} \cup
\mathcal{V}_{\mathrm{I}}^1|_{a_1=\bar{a}_1=0, a_4 \neq 0} \cup
\mathcal{V}_{\mathrm{I}}^1|_{a_1=\bar{a}_1=0, \bar{a}_4 \neq 0} \cup
\mathcal{V}_{\mathrm{I}}^1|_{a_1=\bar{a}_1=a_4=\bar{a}_4=0}, \nn \\
\mathcal{W}&=
\mathcal{W}|_{A_1 \neq 0}
\cup
\mathcal{W}|_{\bar{A}_1 \neq 0}
\cup
\mathcal{W}|_{A_1=\bar{A}_1=0, A_2 \neq 0}
\cup
\mathcal{W}|_{A_1=\bar{A}_1=0, \bar{A}_2 \neq 0}
\cup
\mathcal{W}|_{A_1=\bar{A}_1=A_2=\bar{A}_2=0}, \nn
\end{align}
one has
\begin{align}
\mathcal{V}_{\mathrm{I}}^1=\mathcal{W}, \nn
\end{align}
which proves Prop \ref{point}. \hspace*{\fill} \wsq \\
The algebraic variety $\mathcal{S}$
is isomorphic with the
Segre threefold $\mathbb{P}^1(\mathbb{C}) \times  \mathbb{P}^2(\mathbb{C})$.
\begin{proposition}
\begin{align}
\mathbb{P}^1(\mathbb{C}) \times  \mathbb{P}^2(\mathbb{C}) \cong \mathcal{S}. \nn
\end{align}
\hspace*{\fill} \sq 
\end{proposition}
The map $\psi$
\begin{align}
&\psi:\mathbb{P}^1(\mathbb{C}) \times  \mathbb{P}^2(\mathbb{C}) \longrightarrow 
\mathcal{S}, \nn \\
&\psi((D_1, D_2) \times (E_1, E_2, E_3) )=(D_1 E_1, D_2 E_3, D_1 E_2, 
D_2 E_1, D_1 E_3, D_2 E_2), \label{shazou}
\end{align}
parametrizes the points in 
$\mathcal{S}$ by
$\mathbb{P}^1(\mathbb{C}) \times  \mathbb{P}^2(\mathbb{C})$.
Combining this map with Prop \ref{point}, one sees that
the points in
$\mathcal{V}_{\mathrm{I}}^1$
can be parametrized by the projective variety 
$\mathcal{U}_{\mathrm{I}}=\mathbb{P}^1(\mathbb{C}) \times
\mathbb{P}^1(\mathbb{C}) \times
\mathbb{P}^2(\mathbb{C})$ (Def \ref{algebraicvariety}).
\begin{proposition}
Any point in $\mathcal{V}_{\mathrm{I}}^1$ can be parametrized by
$\mathcal{U}_{\mathrm{I}}$ as
\begin{align}
\mathcal{V}_{\mathrm{I}}^1&=\left\{
\left(
\begin{array}{ccccc}
D_1 D_2 E_1^2 & D_1^2 E_1^2 & D_1 E_2 B_2 & D_1 E_2 B_1 & D_2^2 E_1^2 \\
D_1 D_2 E_3^2 & D_2^2 E_3^2 & D_2 E_2 B_1 & D_2 E_2 B_2 & D_1^2 E_3^2 
\end{array}
\right) \ \Big| \
\begin{array}{l}
(B_1, B_2) \times (D_1, D_2) \times (E_1, E_2, E_3) \\
\in \mathcal{U}_{\mathrm{I}}
\end{array}
\right\}. \nn
\end{align}
\hspace*{\fill} \sq
\label{parametrizationofvariety}
\end{proposition}
Let us denote the solution space corresponding to
$\mathcal{V}_{\mathrm{I}}^0$ and $\mathcal{V}_{\mathrm{I}}^1$ as
$\mathcal{A}_{\mathrm{I}}^0$ and $\mathcal{A}_{\mathrm{I}}^1$ respectively.
\begin{definition}
\begin{align}
\mathcal{A}_{\mathrm{I}}^0&=\{ K_{\mathrm{I}}(z, \mathcal{Q}_{\mathrm{I}})
\ | \ \mathcal{Q}_{\mathrm{I}} \in \mathcal{V}_{\mathrm{I}}^0 \}, \nn \\
\mathcal{A}_{\mathrm{I}}^1&=\{ K_{\mathrm{I}}(z, \mathcal{Q}_{\mathrm{I}})
\ | \ \mathcal{Q}_{\mathrm{I}} \in \mathcal{V}_{\mathrm{I}}^1 \}. \nn
\end{align}
\hspace*{\fill} \sq
\end{definition}
Noting that $\mathcal{A}_{\mathrm{I}}^0$ is
included in $\mathcal{A}_{\mathrm{I}}^1$ as
\begin{align}
\mathcal{A}_{\mathrm{I}}^0=\{K(z)= \textrm{Id} \}
=A_{\mathrm{I}}^1|_{D_2=E_1=E_2=0}, \nn
\end{align}
and utilizing Prop \ref{parametrizationofvariety}, one has
\begin{align}
\mathcal{A}_{\mathrm{I}}=\mathcal{A}_{\mathrm{I}}^0 \cup
\mathcal{A}_{\mathrm{I}}^1
=
\mathcal{A}_{\mathrm{I}}^1=
\{ K_{\mathrm{I}}(z, \mathcal{P}_{\mathrm{I}})
\ | \ \mathcal{P}_{\mathrm{I}} \in \mathcal{U}_{\mathrm{I}} \}
=\mathcal{B}_{\mathrm{I}}, \nn
\end{align}
which proves Prop \ref{spaceidentification} (i).

\hspace*{\fill} \wsq

%
%

\end{document}